\documentclass[10pt,a4paper]{article}
\usepackage[latin1]{inputenc}
\usepackage{amsmath}
\usepackage{amsfonts}
\usepackage{amssymb}
\usepackage{graphicx}
\usepackage{hyperref}
\title{
	Physics, Determinism, and the Brain
}

\author{George F R Ellis}%
\begin{document}
	\maketitle
\begin{abstract}
\textit{This paper responds to claims that causal closure of the underlying microphysics determines brain outcomes as a matter of principle, even if we cannot hope to ever carry out the needed calculations in practice. Following two papers of mine where I claim firstly that downward causation enables  genuine causal powers to occur at higher emergent levels in biology (and hence in the brain) \cite{Ellis_2020a_Emerge}, and that  secondly causal closure is in reality an interlevel affair involving even social levels \cite{Ellis_2020b_Closure}, Carlo Rovelli has engaged with me in a dialogue where he forcefully restates the reductionist position that microphysics alone determines all, specifically the functioning of the brain. Here I respond to that claim in depth, claiming that if one firstly takes into account the difference between synchronic and diachronic emergence, and secondly takes seriously the well established nature of biology in general and neuroscience in particular, my position is indeed correct.}
\end{abstract} 
\tableofcontents

\section{Physics, The Brain, and Predictability}\label{sec:Intro}
After I published articles in \textit{Foundation of Physics}   on emergence in solid state physics and biology \cite{Ellis_2020a_Emerge} and on causal closure in biology and engineering contexts \cite{Ellis_2020b_Closure}, I have had an interesting exchange with Carlo Rovelli questioning my arguments.  This paper responds as regards brain function. 
In Section \ref{sec:dialogue} I will give an edited version of that interchange. In section \ref{sec:response} I outline my response. The key point is that while his argument does indeed apply in the case of synchronic emergence - emergence at one point in time -  it does not apply in the case of diachronic emergence, that is, emergence as it unfolds over time. The  paper is summarised in Section \ref{sec:intro_summary}.

\subsection{A dialogue}\label{sec:dialogue}
Here is a summary of that dialogue. 

\begin{description}
	\item[CR]    
- You list a large number of situations in which we understand phenomena, cause and effects, and we make sense of reality, using high level concepts.  I think this is great. 

- You point out that nobody is able to account for these phenomena in terms of microphysics.  I think this is true.

-  You emphasize the fact that we *need* these high level notions to understand the world.  I think this is important and is an observation which is underappreciated by many.   I agree.

- You point out that to some extent similar points have been made by a number of biologists, solid state physicists, etcetera.   I definitely think this is true.

$\Rightarrow$ Then, you jump to a conclusion about microphysics, which does not follow from any of the above.  And I think that the large majority of physicists also think that it does not follow.  
 Given a phenomenon (condensed matter, biology, anything else), happening in a specific situation in a specific context, it is the belief of the near majority of physicists today that the initial microphysical data (which we may not know), determines uniquely the probability distribution of the later outcomes.   

How does this square with all your examples, and why there is no contradiction between all your examples and the points you make, and this?  The reason is that in all your examples the high-level cause is also a microphysics configuration.   For instance, a crystal configuration determines the motion of the electrons, smoking causes cancer, my getting excited changes the motion of the electrons in my body.   All this is true of course.  But a crystal configuration, smoking, my getting excited are also microphysical states of affairs.  Once we fold them in the microphysics, there is no reason whatsoever for the microphysics not to be sufficient to determine what happens.  There isn't something ``in addition'' to the microphysics.   It is only that the microphysics  is far more complicated than what we can manage directly in our calculations of simple minded understanding. 

Where is our disagreement?   It is not big.  But I think it is crucial, and I think it can be summarized in the case of the effect of Jupiter and the drop of water on the beach of Marseille affected by its gravitational pull.   Here we can take the microphysics to be the Newtonian gravitation of all the small grains of matter in the Solar system, governed by classical mechanics with Newtonian interactions and short scale pressure.  This is a good level of approximation. 
 Can we compute the motion of the drop of water using the Newton equations and all the forces?  Of course we cannot.  There are far too many grains of matter in the Solar system.   But we can go to a higher level description, where we ignore a huge amount of details and we represent everything in terms of a few planets in their orbits.  "Planet" is not a notion in the microphysics.  It is a simple calculation to account for the tidal forces due to the Moon and the Sun and to estimate the small correction due to Jupiter. And we find that the drop of water last Tuesday came a bit higher than expected because of Jupiter. 

Here is the point: there is nothing I see, in all the examples you mention that distinguishes them from this version of``top-down-causation'': The high-level effect of Jupiter affects the motion of a drop of water.   I do not see any difference between this and the crystalline structure affecting motion of electrons in condensed matter, or a biological molecule reacting to an evolutionary pressure, or my excitement affecting the motion of an electron. 


\item[GE] You are then agreeing that high level causation is real? 

\item[CR] "Causal" mean all sort of different things for all sort of different people.  I am not being pedantic, I think it is one of the cruxes of the matter.
Instinctively, I am with Russell that notices that there are no "causes" in physics: there are just regularities expressed by laws. 
But of course I am aware (with Cartwright ) that we do use causes heavily and effectively .  I think that "causes" make sense with respect to an agent that can act, and that the agency of the agent is ultimately rooted in entropy growth.    
If "causal" concepts are understood in this high level and sort of non-fundamental manner, then suddenly your entire project makes sense to me I do not think that anybody would deny that smoking causes cancer.   Therefore, yes, I agree that high level causation is real.
I think that "smoking" denotes a large ensemble of microphysical configurations, all of which (actually, in this context: many of which), evolving according to microphysical laws without any reference to higher order notions, evolve into later microphysical configurations that belong to the ensemble that we call "a person with cancer".

\item[GE] 
``The reason is that in all your examples the  high-level cause is also a microphysics configuration" Yes indeed This sounds very close to 
Denis Noble's Principle of Biological relativity, extended to include the physics level, which is what I propose in my two papers \cite{Ellis_2020a_Emerge} \cite{Ellis_2020b_Closure}. I certainly state that the micro levels are effective, as you do - there is no disagreement there. Maybe then we are not so far apart. 
Question: would you agree that you can invert that statement to get: "in all these examples the  microphysics configuration is also a high-level cause"? 

\item[CR] Here we are back to the ambiguity of the notion of ``cause''.  I do not know what you mean.   The microphysical configurations belonging to ``smoking'' do evolve into those belonging to ``cancer'', while the microphysical configurations ``almost everything the same but not smoking'', do not.    If this is what we mean by ``cause'', I agree.    If there is some other meaning of ``cause'', I sincerely do not understand it.  I think that this is what we more more less mean usually, (what else means ``smoking causes cancer'' if not that by not not smoking we can get less cancer?), then I am happy to use ``cause'' here. 

\item[GE] This is  the key phrase: "Once we fold them in the microphysics". This sounds very much like what I am saying.  The higher level situation is setting the context for the microphysics to act, and hence shaping the specific outcomes that occur through the microphysics If that is what you mean, then we agree! Otherwise what does that phrase mean? 

\item[CR] No, here we disagree, I think.   Because there isn't the microphysics, and then something else that sets the context.   The macro-physics is just a way of talking about microphysical states.  "Smoking" is not something that that is added to the microphysics: it is that one possible configuration of the microphysics.  More precisely: an ensemble of many possible configurations of the microphysics.   
This is the main point: you cannot have the same microphysics with different "contexts".   Different "contexts" always require different microphysics. 
It is the key question: the question I asked you when we started talking about that at a conference somewhere: suppose there are two Chess games on two different planets: everything looks the same but the rules of Chess are a bit different on the two planets (in one you can castle after having moved the king, in the other you cannot). Here a very high order difference (different rules of Chess) make the evolution different in the two planets. 
BUT (here is the point): could the rules in the two planets be exactly the same if the microphysics was the same?  NO, of course, because different rules of the games means different memories stored in the players brains, hence different synapses, hence different physics.   To get a difference, you need different microphysical configurations.  You cannot have a high level source of difference without having a microphysics level difference that achieves the same result. 

\item[GE] 
To get full clarity, please clarify two questions:  

1. Consider one person and the issue of smoking; or (for your neuroscientist friends) one person's brain states when they play chess.  Are you envisaging the total cause of their future behaviour being the microphysics configuration of their own body/brain, or of some larger ensemble of particles? If so, what set? What set does ``causal closure"'', in your terms, refer to?

\item[CR] A larger ensemble of course, because what happens to a person depends on plenty of exterior influences.

\item[GE] The second question.

2. Do you consider ``folding them in the microphysics'' (first reply above) as an ongoing process that is taking place all the time, or not? 

\item[CR] ``Folding then in the microphysics'' means recognizing what they are, once translated into microphysics. They are ensembles of microphysical states.   If you do this at the start of a time interval and if you include enough degrees of freedom to account for anything that matters, then the microscopic evolution does the job, whether or not we can compute. 
If this was NOT the case, we would have found cases whether these evolution laws fail. When people say that a theory is "causally closed" they mean that the initial conditions determine the following (of course: in principle. In practice we cannot do the calculations, nor know all relevant initial conditions.) 

\item[GE] To be clear, is it a process that takes place once at the start of some interaction (the brain starts considering a chess problem) and then the physics by itself determines all outcomes; or is it an ongoing process that is taking place all the time ?

\item[CR]
It is not a process.   I can in principle describe the set of events without any reference to high level concepts, or I can describe the set of events using high level concepts.  Both work.   The difference is that one may be unmanageable and the other may be manageable.    We have these different levels of description everywhere in life: I can think of my trip from Verona to Marseille accounting for the instantaneous changes of velocities, or I can represent it just as average constant velocity with some snack-pauses.   The first is unmanageable, the second uses high level concepts, is more useful.   But it has less information than the first, not more. 

\item[GE] Finally, of course I believe the outcomes lies in the space of possibilities allowed by the microphysics. It could not be otherwise, and the marvel is that that set of laws is able to produce such complexity. There is no way I am underestimating what the physics allows. In the case of biology, I see it as happening because the  physics allows the existence of the Platonic space of all possible proteins discussed by Andreas Wagner in his marvelous book \textit{Arrival of the Fittest} \cite{Wagner 2014}.  

\item[CR] Yes, this is what I meant.  I am glad we agree here.   This space of possibility is immense, and too hard to explore theoretically even if we know the microscopic laws and the fact these laws are not violated and they determine (probabilistic) evolution unequivocally.   Still, we carve it out for our understanding by recognising high level patterns and using them. 
 But then why do you need the microphysics to be affected by something outside it? You do not need it. And we have zero evidence for anything like this.   The autonomy of the higher level logic that you keep citing is no argument against the autonomy of the microphysics.   



But you seem to mean more.   Do you mean that they select which dynamical histories are realised and which not? That is: which initial conditions are allowed and which not?
Or they select which quantum outcomes become actual, over and above the quantum probability amplitudes?
Or what else?

It seems to me you confuse the richness of the tools of a physicist, and the complexity of reality, with a statement about lack of causal closure of the microphysics, that does not follow. 

\end{description}
Finally, a last response:
\begin{description}
	 \item[CR] Today the burden of the proof is not on this side.  Is on the opposite side.  Because:

\textbf{(i)}  There is no single phenomenon in the world where microphysics has been proven wrong (in its domain of validity of velocity, energy, size...).

\textbf{(ii)}  By induction and Occam razor, is a good assumption that in its domain validity it holds.

\textbf{(iii)} There are phenomena too complex to calculate explicitly with microphysics.  These provide no evidence against (ii), they only testify to our limited tools.	 
\end{description}
There is much more, but I will not repeat it all. Rather I summarise points of agreement, of misunderstanding, and of disagreement, and then take up that challenge.

\paragraph{Points of agreement}
\begin{itemize}
	\item All is based in immutable lower level physical laws, unaffected by context
	\item  These laws allow immensely complex outcomes when applied to very complex microstates, such as those that underlie a human brain
	\item Testable higher level laws correctly express dynamics at higher levels 
	\item They make processes at that level transparent in a way that is completely hidden when one traces that same dynamics at the lower levels
\end{itemize}

\paragraph{Point of misunderstanding}

\begin{itemize}
	\item \textbf{Downward causation}
	
	[\textbf{CR}] From your examples it does not follow that our current elementary theories, such GR and the SM, have free 
	parameters that are controlled by something else that we understand.  That would be a wild speculation unsupported by anything.
	
[\textbf{GE}] The theories themselves of course do not have such parameters, and I have never claimed that they do. But Lagrangians used in specific contexts do. 
	
	Those theories  \textit{per se } have generic Lagrangians that apply to anything at all, and so say nothing detailed  about anything specific. They do not by themselves determine outcomes in biology or engineering. A particular context determines the details of the terms in the Lagrangian, and that happens in a contextually dependent way: after all, in a particular case it represents a specific context. Once the Lagrangian has been determined at time $t_1$ then the next emergent step is indeed fully determined purely at the micro physical level, as Carlo claims. But macro conditions can then change  parameters  in the Lagrangian. That is where the downward causal effects come in.  
	For specific examples,  the case of transistors in digital computers is discussed in \cite{Ellis_Drossel_computers}, and voltage gated ion channels in the brain in  \cite{Ellis_Kopel_2019}.

The key issue is whether downward causation is real, having  real causal powers. I argue that it is; and that this kind of causation does not required any compromising of the underlying physics. It works by changing constraints \cite{Juarrero 2002}.  
\end{itemize}

\paragraph{Points of disagreement}

\begin{description}
	\item[CR] 
	\textbf{(i)} In principle, we could entirely deduce what happens even ignoring the high-level concepts
	
	 \textbf{(ii)} In practice, we use high-level concepts
	
	 \textbf{(iii)} In addition, we get a better grasp, a better sense of "understanding", a better control, in terms of higher level notions.

Hence: high-level notions are far more relevant for comprehension of the world.
I think that (ii) and (iii) are very important.   

(\textbf{iv}) But if you add the negation of (i), you get everybody disagreeing (me included) and the message about (ii) and (iii) does not go through.
	\item [GE] \textbf{1} Causation in biology is an interlevel concept \cite{Mossio 2013}. Physics underlies this but does not by itself give causal closure. What Carlo calls ``causal closure of physics'' is in fact the statement that at its own level, it is a well-posed theory: a completely different affair.
	
	 \textbf{2} Carlo's statement can be defended in the synchronic case (at a time), but is not always true  for individual brains in the diachronic case (unfolding over time). 
	
	 \textbf{3} Statistics don't cut it. Individual events occur. We have to explain why specific individual brain events occur, for example leading to the specific words in Carlo's emails. Specific events in individual lives occur and need to be accounted for.
	 
	 \textbf{4} Microphysics enables this but does not determine the outcomes. 
	 The basic physics interactions of course enable all this to happen: they allow incredible complexity to emerge. 
	 Higher level organising principles such as Darwin's theory of evolution then come into play. That then changes the macro level context and hence the micro level context. This downward process \cite{Campbell_1974} relies on concepts such as `living' that simply cannot be represented at the microlevel, but determine outcomes.
	 
	 \textbf{5} The reductionist physics view is based in a linear view of causation.  Central to the way biology works are the closely related ideas of \textit{self cause} and \textit{circular causation}.
\end{description}
\subsection{A response}\label{sec:response}
In the rest of this paper, I give a full response to Carlo's arguments in the case of the brain, based in the nature of causation in biology. It rests on three things. First, taking seriously the nature of biology in general \cite{Campbell and Reece 2008} and  neuroscience \cite{Kandel 2012} \cite{Kandel and Schwartz} in particular, demanding that whatever overall theory we propose must respect that nature.   Second, requiring that individual events and outcomes are what need to be accounted for, not just statistics. Third, noting the key difference between synchronic and diachronic emergence, which we did not make in our email interchange. The answer is very different in these two cases. 

\paragraph{Synchronic and diachronic emergence} 
 Carlo's argument - the microstate uniquely determines macro level outcomes - is correct when we consider synchronic emergence. That is what a lot of neuroscience is about. It is not valid however when one considers diachronic emergence. 
 The issue here is one of timescales. 

\textit{Synchronic emergence} is when the timescale $\delta t :=t_b-t_a$ of the considered microdynamic outcomes  is very short relative to the timescale $\delta T$ of change  of structures at the micro scale: $\delta T \gg \delta t$.  It is the issue of emergent dynamics when parameters are constant and constraints unchanging. In the case of the brain   this would for example be the flow of electrons in axons leading to mental outcomes at that time, with this micro structure taken as unchanging. Electrons and ions flow in a given set of neural connections. 

\textit{Diachronic  emergence} is when the timescale  of micro dynamic outcomes considered $\delta t$ is of the same order or larger than  the timescale $\delta T$ of change of structure  at the micro scale: $\delta T \leq \delta t$, so microdynamics contexts alters significantly during this time. It is the case when parameters or constraints change because of interactions that are taking place. In the case of the brain this would for example be when something new is learned so that strengths of neural connections are altered.

\paragraph{Consider first a single brain} %
Dynamic outcomes at the molecular  scale are due to the specific structures at the cellular scale, neural connectivity for example, and the way that they in turn constrain electron and ion activity. Three points arise.
\begin{itemize}
	\item First, \textbf{\textit{the brain is an open system}}. It is not possible for the initial physical state to determine  later states because of the flood of data incoming all the time. The last round of  microlevel data does not determine the initial data  that applies at the next round of synchronic emergence. The brain has evolved a set of mechanisms that enable it to cope with the stream of new data flowing in all the time by perceiving its meaning, predicting futures, and planning how to respond. This is what determines outcomes rather than evolution from the last round of initial data.  
	\item Second, \textbf{\textit{the brain is a plastic brain}} that changes over time as it learns. Neural connections are altered as learning takes place in response to the incoming stream of data. This change in constraints alters future patterns of electron and ion flows. This learning involves higher level variables and understandings such as ``A global Coronavirus pandemic is taking place'', that cannot be characterised at lower levels and cannot be predicted from the initial brain microdata.  
	\item Third, \textbf{\textit{there is a great deal of stochasticity at the molecular level}} that breaks the ideal of Laplacian determinism at that level. Molecular machines have been evolved that take advantage of that stochasticity to extract order from chaos. From a higher level perspective, this stochasticity enables organisms to select lower level outcomes that are advantageous to higher level needs. From a systems perspective, this enables higher level organising principles such as existence of dynamical system basins of attraction to determine outcomes. 
\end{itemize} 
This argument applies  to all biology, as all biological systems are by their nature open systems \cite{Peacock 1989}. The initial physics data for any organism by itself cannot in principle determine specific later outcomes because of this openness. 

The fundamental physical laws are not altered or overwritten when this happens; rather the context in which they operate - for example opening or closing of ion channels in axons - determine what the specific outcomes of generic physical laws will be as alter configuration. From a physics viewpoint this is represented by time dependent constraint terms or potentials in the underlying Hamiltonian \cite{Ellis_Kopel_2019}.



\paragraph{The whole universe gambit} \label{sec:intro_whole_universe}
The ultimate physicalist response is ``Yes the brain may be an open system but the whole universe is not; and the brain is just part of the universe, which is causally complete. Hence brain dynamics is controlled by the microphysics alone when one takes this into account, because it determines all the incoming information to the brain''.  However  this argument fails for the following reasons:
 \begin{itemize}
 	\item Firstly there is \textbf{\textit{irreducible quantum uncertainty}} in outcomes, which implies the lower physics levels are in fact not causally complete. This can get  amplified to macroscales by  mechanisms that change mental outcomes, such as altered gene expression due to damage by high energy photons.
 
\item   Secondly, this downward process - inflow of outside information to individual brains - does not uniquely determine brain how microstructures change through memory processes because of 
\textbf{\textit{multiple realisability}}. But such  uniqueness is required to sustain a claim that the causal closedness of microphysics determines specific brain outcomes over time.

 \item  Thirdly,  \textbf{\textit{chaotic dynamics}} associated with strange attractors occurs, which means the emergent dynamics of weather patterns is not in fact predictable even in principle over sufficiently long timescales. This affects decisions such as whether to  take an umbrella when going to the shops or not. 
  
\item Fourthly, \textbf{\textit{microbiome dynamics}} in the external world affects brain outcomes in unpredictable ways, for example when a global pandemic occurs

 \item Fifthly, this all takes place in a social context where \textit{\textbf{social interactions}} take place between many brains, each of which is itself an open system. Irreducible uncertainty influences such contexts  due to the real butterfly effect (weather) and the impossibility, due to the molecular storm, of predicting specific microbiome mutations that occur (e.g. COVID-19), leading to social policy decisions, 
 that are high level variables 
  influencing 
    macro level brain states. The outcomes then influence details of synaptic connections and hence shape future electron and ion flows. 
 
  \end{itemize} 
This downward causation from the social/psychological level to action potential spike chains and synapse activation is essential to the specific outcomes that occur at the physical level of electron and ion flows in individual brains. Causal closure only follows when we include those higher level variables in the dynamics. 

\subsection{The argument that follows} \label{sec:intro_summary}
Section \ref{sec:foundations} sets the scene by discussing the foundations for what follows, in particular the fact that life is an ongoing adaptive process. In the following sections I discuss the key issues that support my view. 

Firstly, an individual brain is an open system, and has been adapted to handle the problems this represents in successfully navigating the world (Section \ref{sec:brain_open_system}). This rather than the initial brain micro  data determines outcomes.

Secondly, the brain learns: it is plastic at both macro and micro levels, which continually changes the context within which the lower level physics operates (Section \ref{sec:Plastic_brain}). 

Third, the kind of Laplacian view of determinism underlying Carlo's position is broken at the molecular level because of the huge degree of stochasticity that happens at that level (Section \ref{sec:stochasticity}). Biological processes - such as Darwinian evolution, action choices, and the brain pursuing a line of logical argumentation  -  are what in fact determine outcomes, taking advantage of that stochasticity. Biological causation occurs selects   preferred  outcomes from the molecular storm, and the brain selects from action options. 


 In section \ref{sec:whole_universe} I counter the whole universe gambit by claiming that this will not work 
 because 
 of quantum wave function collapse,  
macro level chaotic dynamics, multiple realisability of macro brain states, 
and unpredictable microbiome interactions that affect brain dynamics both directly and via their social outcomes. 

Section \ref{sec:conclude} consider how higher level organising principles - the effective laws that operate at higher levels - are in fact what shapes outcomes. This is what enables causal closure - an interlevel affair - in  practice. I also comment on the issue of freewill (Section \ref{sec:FreeWill}). 

\section{Foundations}\label{sec:foundations}
As stated above, the premise of this paper is that when relating physics to life, one should take seriously the nature of biology as well as that of physics. 
I  assume the standard underlying microphysics for everyday life, based in the  Lagrangian for electrons, protons, and nuclei, see \cite{Laughlin_Pines_2000} and \cite{Bishop_2005_patch}.
This section sets the foundation for what follows by discussing the nature of biology and of causation. 

Section {\ref{sec:nature_biology} discusses the basic nature of biology. Section \ref{sec:hierarchy} outlines the biological hierarchy of structure and function.  Section \ref{sec:effective_theories} discusses the nature of Effective Theories at each emergent level \textbf{L}. Section \ref{sec:equal_validity_levels} discusses the equal validity of each level in causal terms. Section \ref{sec:causation_types} discusses the various types of downward causation, and Aristotle's four types of causes as well as Tinbergen's `Why' questions. Section \ref{sec:multiple_realise} discusses the important issue of multiple realisability of higher level structure and function at lower levels. Finally Section \ref{sec:Protectates_HLOP} discusses the key role of Higher Level Organising Principles.
 
\subsection{The basic nature of biology}\label{sec:nature_biology}

All life \cite{Campbell and Reece 2008} is based in the interplay between structure (that is, physiology \cite{Guyton 2016}  \cite{Physiology_human}) and function.
 For good functional, developmental, and evolutionary  reasons, it is composed (\textbf{Table 1}:\S\ref{sec:hierarchy}) of \textbf{\textit{Adaptive Modular Hierarchical Structures}} \cite{Simon 2019} \cite{Booch 2006}  based in the underlying physics. It comes into being via the interaction between evolutionary and developmental (Evo-Devo)
  processes \cite{Carroll 2005} \cite{Carroll 2008}, and has three key aspects.
\footnote{An excellent introduction to the relevant mechanisms is given in  \cite{Noble 2016}.}

\paragraph{1. Teleonomy: function/purpose}
Life has a teleonomic nature, where \href{https://en.wikipedia.org/wiki/Jacques_Monod}{Jacques Monod} defines teleonomy as the characteristic of being "endowed with a purpose or project"  (\cite{Monod_1971}:9)
 He points out the extreme efficiency of the teleonomic apparatus in accomplishing the preservation and reproduction of the structure.
As summarised by  Nobel Prizewinner \href{https://en.wikipedia.org/wiki/Leland_H._Hartwell}{Leland Hartwell} and colleagues \cite{Hartwell_et_al_1999}, 
\begin{quote}
	``\textit{Although living systems obey the laws of
		physics and chemistry, the notion of
		function or purpose differentiates biology
		from other natural sciences. Organisms
		exist to reproduce, whereas, outside religious
		belief, rocks and stars have no purpose.
		Selection for function has produced the living
		cell, with a unique set of properties that
		distinguish it from inanimate systems of
		interacting molecules. Cells exist far from
		thermal equilibrium by harvesting energy
		from their environment. They are composed
		of thousands of different types of molecule.
		They contain information for their survival
		and reproduction, in the form of their DNA}''.
\end{quote}
Function and purpose emerge at the cell level. 
 \href{https://en.wikipedia.org/wiki/Fran%C3%A7ois_Jacob}
 	{Francois Jacob} says \cite{Jacob 1974}\footnote{Quoted in \cite{Peacock 1989}:275.}
\begin{quote}
	\textit{``At each level of organisation novelties appear in both properties and logic. To reproduce is not within the power of any single molecule by itself. This faculty appears only within the power of the simplest integron\footnote{An `Integron' is each of the units in a hierarchy of discontinuous units formed by integration of sub-units of the level below \cite{Jacob 1974}:302.} deserving to be called a living organism, that is, the cell. But thereafter the rules of the game change. At the higher level integron, the cell population, natural selection imposes new constraints and offers new possibilities. In this way, and without ceasing to obey the principles that govern inanimate systems, living systems become subject to phenomena that have no meaning at the lower level. Biology can neither be reduced to physics, nor do without it.}''
\end{quote}

\paragraph{2. Life is a process}
Being alive is not a physical thing made of any specific elements. It is a \textit{process} that occurs at macro levels, in an interconnected way. In the case of human beings it involves all levels\footnote{See \textbf{Table 1}, Section \ref{sec:hierarchy}.} from \textbf{L4} (the cellular level) to Level \textbf{L6} (individual human beings), allowing causal closure \cite{Mossio 2013} \cite{Mossio  and Moreno 2010} and hence self-causation \cite{Juarrero 2002} \cite{Murphy and Brown}. 
\begin{quote}
	\textbf{Life is an ongoing adaptive  process} \textbf{\textit{involving metabolism, homeostasis, defence, and learning in the short term, reproduction, growth, and development in the medium term, and evolution in the long term. It uses energy, disposes of waste heat and products,   and uses contextual information to attain its purposes.}}
\end{quote}
The claim I make is that this process of living has causal power, making things happen in an ongoing way. High level processes take place via an interlevel dialogue between  levels \cite{Noble2008_Music}, higher levels  continually altering  the context of the underlying physical levels 
in order to carry out these functions \cite{Ellis_Kopel_2019}. Yes of course the resulting physical processes can be traced out at the physics level. But my claim will be that biological imperatives \cite{Campbell and Reece 2008} enabled by physiological systems  \cite{Physiology_human} \cite{Guyton 2016} shape what happens. Evolutionary processes \cite{Mayr 2001} \cite{Carroll 2008} have enabled this synergy to occur \cite{Noble 2016}.  

\paragraph{3. Basic biological needs and functions}
In the case of animal life,\footnote{Other forms of life share \textbf{B1-\textbf{3}.}} the basic biological functions are,  

\textbf{B1:} Metabolism (acquiring energy and matter, getting rid of waste), 

\textbf{B2:} Homeostasis and defence, 

\textbf{B3:} Reproduction and subsequent development, 

\textbf{B4:} Mobility and the ability to act, 

\textbf{B5:} Information acquisition and processing. 

\noindent They serve as attractors when variation takes places (\cite{Ginsburg and Jablonka 2019}:245). They are the higher level organising principles that evolution discovers and then embodies in hierarchically structured  physiological systems, where the macro functions are supported at the micro level by metabolic networks, gene regulatory networks, and cell signalling networks,  selected from an abstract space of possibilities and realised through specific proteins \cite{Wagner 2014}. 
Information is central to what happens \cite{Nurse 2008} \cite{Davies 2019}.
	
	These principles cannot be described or identified at the underlying microphysical levels not just because the relevant variables are not available at that level, but because their multiple realisability at lower levels means they do not correspond to specific patterns of interactions at the ion and electron level. They correspond to a whole equivalence class of such patterns of interactions (Section \ref{sec:multiple_realise}). 
	
\paragraph{4. Interaction networks} These processes are realised by means of immensely complex \textit{interaction networks} at the molecular level 
\cite{Buchanan 2010} \cite{Junker and Schreiber 2011}:

\textbf{N1}: Metabolic Networks (\cite{Wagner 2014} \S3) \cite{Noble 2016}

\textbf{N2:} Gene Regulatory Networks (\cite{Wagner 2014} \S5) 

\textbf{N3:} Signalling Networks  \cite{Junker and Schreiber 2011} \cite{Buchanan 2010} 

\textbf{N4:} Protein Interaction Networks \cite{Junker and Schreiber 2011}

\noindent based in very complex molecular interactions \cite{Berridge} and with higher level design principles shaping their structure \cite{Alon 2006}, and at the cellular level, 

\textbf{N5:} Neural Networks \cite{Kandel and Schwartz} \cite{Churchland  and  Sejnowski}

\noindent These networks compute in the sense of (\cite{Churchland  and  Sejnowski}:69-74)

\paragraph{5. Branching causal logic}
In order to meet these needs, the dynamics followed 
at each level of biological hierarchies 
is based on contextually informed  dynamical branching  $L$ that 
support the functions $\alpha$ of a trait $T$   in a specific environmental context $E$ \cite{Ellis_Kopel_2019}. 
Thus biological dynamics can be 
functionally-directed rather than driven by inevitability or chance: 
\begin{eqnarray}\label{eg:111}
\texttt 
{Biological dynamics tends to further the function } \alpha\nonumber \texttt{ of a trait } T \\\texttt{ through 
	contextually informed branching dynamics  } L
\end{eqnarray}
where the dynamics $L$ 
in its simplest form   is branching logic 
of the form \cite{Hoffman}
\begin{equation}\label{eq:2}
\texttt{ 
	L: given context }\, C,\,\texttt{ IF}\,\, T(\textbf{X})\,\, \texttt{THEN}\, F1(\textbf{Y}),\,\texttt{ ELSE}\,\,\textbf{} F2(\textbf{Z})
\end{equation}
(a default unstated ``ELSE'' is always to leave the status quo). Here $\textbf{X}$ is a contextual variable which can have many dimensions,    
$\textbf{Y}$ and $\textbf{Z}$ are variables that may be the same variables as \textbf{X} or not.  $T(\textbf{X})$ is the truth value of arbitrary evaluative statements depending on \textbf{X}. It can be any combination of Boolean logical operations (NOT, AND, OR, NOR, etc.) and mathematical operations, while  $F1(\textbf{Y}) $ and $F2(\textbf{Z})$ are 
outcomes tending to further the function $\alpha$.  Thus they might be the homeostatic response ``If blood sugar levels are too high, release insulin'',  or the conscious dynamic 
``If the weather forecast says it will rain, take an umbrella''. 
At the molecular level, these operations are based in the lock and key molecular recognition mechanism (\cite{Noble 2016}:71), \cite{Berridge}.  This mechanism is how information \cite{Nurse 2008} \cite{Davies 2019} gets to shape physical outcomes. 

\paragraph{6. Brain Function} The human brain supports all these activities by a series of higher level processes and functions. These are \cite{Purves et al} \cite{Gray}

\textbf{BR1:} Sensation, perception, classification

\textbf{BR2:} Prediction, planning, making decisions, and action

\textbf{BR3:} Experimenting, learning, and remembering

\textbf{BR4:} Experiencing and responding to emotions 

\textbf{BR5:} Interacting socially, communicating by symbols and  language

\textbf{BR6:} Metacognition, analysis, and reflection, `off-line' exploration of possibilities.

\noindent It does so via its complex adaptive modular hierarchical structure  \cite{Kandel and Schwartz} \cite{Scot 2002}. Brains compute \cite{Marr 2010} \cite{Churchland  and  Sejnowski}, but they are not digital computers \cite{Piccinini and  Shagrir 2014}.

\subsection{The hierarchy}\label{sec:hierarchy}
\noindent The framework for the following is the 
hierarchy of structure and function for the biological sciences shown in (\textbf{Table 1}), based in the underlying physics.

\vspace{0.1in}
\begin{tabular}{|c|c|c|c|}
	\hline \hline
	&  \textbf{Biology Levels} & \textbf{Processes} \\	
	\hline Level 8 (\textbf{L8}) & Environment &  Ecological,  environmental processes  \\ 
	\hline 
	Level 7	(\textbf{L7}) & Society  &  Social processes \\ 
	\hline 
	Level 6	(\textbf{L6})  &  Individuals & Psychological processes, actions \\ 
	\hline 
	Level 5	(\textbf{L5}) &  Physiological systems & Homeostasis, emergent functions \\ 
	\hline 
	Level 4	(\textbf{L4}) & Cells &  Basic processes of life \\ 
	\hline 
	Level 3 (\textbf{L3}) & Biomolecules & Gene regulation, metabolism \\ 
	\hline 
	Level 2 (\textbf{L2}) & Atom, ion, electron Physics & Atomic, ionic, electron interactions
	\\ 
	\hline 
	Level 1	(\textbf{L1}) &  Particle and Nuclear Physics & Quark, lepton interactions  \\ 
	\hline \hline
\end{tabular} 
\vspace{0.1in}

\noindent \textbf{Table 1}: \textit{The hierarchy of structure  for biology (left) and corresponding processes (right).} \textbf{L2} \textit{is the relevant physics level of emergence,} \textbf{L4} \textit{the fundamental biological level, made possible by} \textbf{L3}  \textit{(in particular proteins, RNA, DNA), in turn made possible by} \textbf{L2} and so \textbf{L1}.\\

The first level where the processes of life occur is \textbf{L4}, the level of cells. At level \textbf{L6} one finds the integrated processes of an individual organism. At level \textbf{L7} one finds sociology, economics, politics, and legal systems.

\subsection{Effective theories}\label{sec:effective_theories}
 I am assuming that each of these levels exists as a matter of fact - they exist ontologically. The key issue is, if we propose a specific level \textbf{L} exists ontologically, there should be a valid Effective Theory $\textbf{ET}_\textbf{L}$ applicable at that level which characterizes that level.  `Valid' means it either makes testable predictions that have been confirmed, or at least characterizes the variables that would enter such a relation.\footnote{The cautionary note reflects the difficulty in establishing reliable relations at levels \textbf{L6}-\textbf{L8}. The theories may have to be described in terms of propensities rather than mathematical  laws. They are nevertheless well established fields of study, for example \cite{Gray} at Level \textbf{L6}, \cite{Berger} at Level \textbf{L7},  and \cite{Houghton 2009} at Level \textbf{L8}.}   Here following \cite{Ellis_2020a_Emerge} and \cite{Ellis_2020b_Closure}, 
one can characterise an Effective Theory $\textbf{ET}_\textbf{L}(a_\textbf{L})$ valid at some level \textbf{L} as follows:
\begin{quote}
	\textit{An} \textbf{Effective Theory} $\textbf{ET}_\textbf{L}(a_\textbf{L})$ \textit{at
		an emergent  level \textbf{L} is a reliable relation between initial conditions described by effective variables $v_\textbf{L} \in \textbf{L}$ and outcomes $o_\textbf{L}\in \textbf{L}$:}
	\begin{equation}\label{eq:effective_laws}
	\textbf{ET}_\textbf{L}(a_\textbf{L}): v_\textbf{L} \in \textbf{L} \rightarrow \textbf{ET}_\textbf{L}(a_\textbf{L})[v_\textbf{L}] = o_\textbf{L} \in \textbf{L}
	\end{equation}
	\textit{where $a_\textbf{L}$ are parameters of the relation, and  $\textbf{ET}_\textbf{L}(a_\textbf{L})$ may be an exact or statistical law. The parameters $a_\textbf{L}$ may be vectorial or tensorial
}\end{quote}   
Thus I will define a meaningful level to exist if there is such a relation. Determining that relation is in effect epistemology, but what it indicates is the underlying ontology.  

The effective theory $\textbf{ET}_\textbf{L}(a_\textbf{L})$ is \textbf{well posed} if for specific choices of the parameters $a_\textbf{L}$ it provides a unique mapping 
(\ref{eq:effective_laws}) from $v_\textbf{L}$ to $o_\textbf{L}$. This is the concept one should use instead or referring to the theory as being causally complete. That is a misnomer because firstly, the idea of causality does not apply to the physics laws \textit{per se} (although effective theories do), and secondly because causal completion - the set of conditions that actually determine what outcomes will occur in real-world contexts - is always an interlevel affair,
no single level \textbf{L} by itself is causally complete (Section \ref{sec:causal_closure}). Effective Theories  represent verifiable patterns of causation at the relevant level, not causal closure \cite{Ellis_2020b_Closure}. 

\paragraph{Effective theory examples}
It is useful to give some examples of effective theories at different levels. It is my contention, in agreement with \cite{Noble 2012} \cite{Noble 2016}, that  real causal processes are going on at each of these levels, even though this is enabled by underlying levels, including the physics ones. The relevant effective theories  are more than just useful descriptions of high level processes. In all but the last two cases this is demonstrated by the fact that evolution has selected genomes that result in them happening. Their causal effectiveness is a driver of evolutionary selection.

\begin{description}

\item[1. Gene regulation] The kind of gene regulatory processes discovered by Jacob and Monod \cite{Jacob and Monod 1961} \cite{Monod_1971} represent real causal processes at the cellular level (they require the relevant molecular processes, but can only take place in the context of a functioning cell \cite{Hofmeyer 2018 Constructors}). Their importance is that they underlie the Evo-Devo processes discussed in \cite{Carroll 2005} \cite{Carroll 2008}. 

\item[2. Action potential propagation] Brain processes are supported at the micro level by propagation of action potential spikes according to the Hodgkin-Huxley Equations
\cite{Hodgkin_Huxley}. This is an emergent phenomenon that cannot be deduced from the underlying physics \textit{per se} because they involve constants that are not fundamental physical constants.    
\cite{Woodward 2018} defends the view that the explanation the equations provide are causal in the
interventionist sense.

\item[3. The brain] The way the brain works overall  \cite{Kandel 2012} \cite{Gray} is based in the underlying neuroscience \cite{Kandel and Schwartz}. It has been arrived at by an evolutionary process based in the advantages its specific functioning provides. Two key issues are the ability to function under uncertainty  \cite{Clark 2013} \cite{Clark 2016} \cite{Hohwy 2013} \cite{Hohwy 2016} and the existence of a symbolic ability  \cite{Deacon 1997} that allows language,  culture, and technology  to arise \cite{Ginsburg and Jablonka 2019}. 

\item[4. Natural Selection] Natural selection  \cite{Mayr 2001} is a meta-principle: it is a process of downward causation  \cite{Campbell_1974} that allows the others listed above to come into being. Because the biological needs listed above are attractor states in the adaptive landscape \cite{McGhee 2011}, evolutionary convergence takes place  \cite{McGhee 2006}: that is, there are multiple ways they can be met. Any physiological implementation in the equivalence class that satisfies the need will do. Thus this is an example of multiple realisability (Section \ref{sec:multiple_realise}), which characterizes topdown causation \cite{Ellis_2016}.   

\item[5. Smoking, lung cancer, and death] The relation  between smoking and lung cancer is an established causal link, as discussed in depth in \cite{Pearl_Why}. It can certainly be redescribed at the physics level, but the key concepts in the correlation - smoking, cancer - cannot.  Therefore, starting off with an initial state described at the microphysics level, one cannot even in principle determine the probabilities of cancer occurring on the basis of those variables alone, let alone when death will occur as a result of the cancer, because death also cannot be described at that level. 

Once cancer occurs (at the genetic/cellular levels \textbf{L3/L4}) leading to  death (at the whole organism level \textbf{L6}) this will alter physical outcomes at the ion/electron level \textbf{L2} because the process of life (see above) has ceased. This is a real causal chain, not just a handy redescription of micro physics: smoking causes cancer and then death as a matter of fact. The physics allows this of course, but the actual physical trajectories and outcomes follows from the essential higher level dynamics of the cessation of being alive. 
\end{description}

\subsection{Equal Validity of Levels}\label{sec:equal_validity_levels}
There is a valid Effective Theory $\textbf{ET}_\textbf{L}$ at each level \textbf{L}, each of them represents a causally valid theory  holding at its level, none more fundamental than the others.   This is expressed nicely in \cite{Schweber 1993}, commenting on Phil Anderson's views:
\begin{quote}
	\textit{``Anderson  believes in  emergent laws.   He holds the
		view that each level has its  own ``fundamental'' laws and its own ontology. Translated into the language of particle physicists, Anderson would say each level has its effective Lagrangian and its set of quasistable particles. In  each level the effective Lagrangian - the ``fundamental'' description at that level -  is  the best  we can  do.'' 
	}
\end{quote}
None of them can be deemed to be more fundamental than any other, \textit{inter alia} because none of them is  \textit{the} fundamental level}, i.e. none is the hoped for Theory of Everything (TOE).
This has to be the case because we don't know the underlying TOE, if there is one, and so don't - and can't - use it in real applications. So all the physics laws we use in applications are effective  theories in the sense of \cite{Castellani_02}, applicable at the appropriate level. 
Similarly, there are very well tested effective theories at levels \textbf{L3-L5} in biology: the molecular level, the cellular level, the physiological systems level for example. Whenever there are well established laws at the higher levels (for example the laws of perception at Level \textbf{L6}) the same applies to them too.

More fundamentally, 
this equal causal validity occurs because  higher levels are linked to lower levels by a combination of upwards and downwards  causation \cite{Noble 2012} \cite{Noble 2016} so no level by itself is causally complete. They interact with each other with each level playing a role in causal completeness. Hence (\cite{Noble 2016}:160),
\begin{quote}
	\textit{\textbf{The Principle of Biological Relativity}: There is no privileged level of causation in biology: living organisms are multi-level open stochastic systems in which the behaviour at any level depends on higher and lower levels and cannot be fully understood in isolation}
\end{quote}
This is because of circular causality which for example necessarily involves downward causation from the whole cell to influence the behaviour of its molecules just as much as upward causation form the molecular level to the cellular level \cite{Noble 2016}:163-164). This applies to all levels in \textbf{Table 1}, i.e. it includes the underlying physics levels as well \cite{Ellis_Kopel_2019} 
\cite{Ellis_2020b_Closure}, as has to be the case for physical consistency. 

In the case of the brain, after having set out in depth the hierarchical structure of the brain (\cite{Churchland  and  Sejnowski}:11,27-48), Churchland  and  Sejnowski state (\cite{Churchland  and  Sejnowski}:415) 
\begin{quote}
	``\textit{An explanation of higher level phenomena in terms of lower level phenomena is usually referred to as a} reduction, \textit{though not in the perjorative sense that implies the higher levels are unreal, explanatorily dismissable, or somehow empirically dubious'', }
\end{quote}
which agrees with the view put here. Brain computational processes have real causal power \cite{Marr 2010} \cite{Scot 2002} \cite{Churchland  and  Sejnowski}.

\subsection{Types of causation}\label{sec:causation_types}
Causation can be characterised either in an interventionist or a counterfactual sense, either indicating when causation takes place \cite{Pearl 2009} \cite{Pearl_Why}. The first key claim I make is that as well as upward causation, downward causation takes place \cite{Noble 2012} \cite{Ellis_2016}. The second one is that as well as efficient causation, Aristotle's other forms of causation play a key role in real world outcomes. 

\paragraph{Downward causation}
Physicists take for granted upward causation, leading to emergence through aggregation effects such as coarse graining. However one can claim there is also downward causation that occurs via various mechanisms \cite{Noble2008_Music} \cite{Ellis 2012} \cite{Ellis_2016}, allowing strong emergence \cite{Chalmers_2000} to occur. Carlo agrees downward causation takes place, but believes it can be rewritten purely in terms of low level physics, and hence does not represent strong emergence.

Downwards effects in a biological system occur 
because of physiological processes \cite{Noble2008_Music}, \cite{Noble 2012}. These processes \cite{Guyton 2016} are mediated at the molecular level by developmental systems \cite{Oyama Griffiths Cycles} operating through metabolic and 
gene regulator networks \cite{Wagner 2014} and cell signalling networks \cite{Berridge}, guided by higher level physiological needs. They reach down to the underlying physical level \textbf{L2} via time dependent constraints 
\cite{Ellis_Kopel_2019}. The \textit{set of interactions} between elements at that level is uniquely characterised by the laws of physics ${L}$, but their \textit{specific outcomes} are determined by the biological context in which they operate.\\
An example is 
determination of    \href{https://en.wikipedia.org/wiki/Heart#Heart_rate}{heart rate}. Pacemaker activity of the heart is via cells in the sinoatrial node that create an action potential and so alter ion channel outcomes. This pacemaking circuit is an integrative characteristic of the system as a whole \cite{Fink_and_Noble} -  that is, it is an essentially higher level variable - that acts down to the molecular level \cite{Noble 2012} \cite{Noble 2016}. 
In the synchronic case - nothing changes at either macro or micro levels - it is correct that one can predict the lower level and hence the higher level dynamics purely from the lower level initial state. However if the higher level state changes - an athlete starts running - the higher level state changes, and this alters lower level conditions. Nothing about the initial molecular level state of the heart or the underlying physics state could predict this happening. Neither could initial knowledge of both the athletes heart and brain micro states determine this outcome, because it depended on an external event - the firing of the starting gun, another macro level event which the athlete's initial  states cannot determine. 

Considering the individual athlete, causation at the macro level is real: the firing of the starting gun led to her leaving the starting post. Downward causation that alters motion of ATP molecules in her muscles via metabolic networks is real: that is a well established physiological process \cite{Physiology_human}. The result is altered electron flows in the muscles, in a way consistent with the laws of physics but unpredictable from her initial microphysical state. Regression to include the brain state of the person firing the gun will not save the situation, as one then has to include all the influences on his brain state \cite{Nobleetal2019} as well as all the stochastic elements in his brain (Section \ref{sec:stochastic_brain}). 

A similar example of a rhythmic pattern determined by a network as a whole is the stomatogastric ganglion of the spiny lobster (\cite{Churchland  and  Sejnowski}:4-5): 
\begin{quote}
	``\textit{The network in question contains about 28 neurons and serves to drive the muscles controlling the teeth of the gastric mill so that food can be ground up for digestion. The output of the network is rhythmic, and hence the muscular action and the grinders movements are correspondingly rhythmic. The basic electrophysiological and anatomical features of the neurons have been catalogued, so that the microlevel vitae for each cell in the network is impressively detailed. What is not understood is how the cells in the network interact to constitute a circuit that produces the rhythmic pattern. No one cell is a repository for the cells rhythmic output; no one cell is itself the repository for the properties displayed by the network as a whole. Where then does the rhythmicity come from? Very roughly speaking, from the patterns of interactions among cells} and \textit{the intrinsic properties of the component cells.}
\end{quote}
The network produces rhythmic patterns in the cells, which produce rhythmic activity in the constitutive electrons and ions. This is a classic example of higher level order controlling both macro and micro level outcomes.

\paragraph{Types of downward causation}
The basic type of downward causation are as follows (developed from \cite{Ellis 2012} \cite{Noble 2012} \cite{Noble 2016} \cite{Ellis_2016}):
\begin{description}
	\item [TD1A] \textbf{Boundary conditions} are constraints on particles in a system arising from the environment\footnote{Carlo's example of Jupiter causing tides on Earth fits here: Jupiter is part of the Earth's environment, causing a detectable gravitational field at Marseilles.} 
	as in the case of a cylinder determining pressure and temperature of the enclosed gas, or the shape of tongue and lips determining air vibrations and so spoken words. 
	\textbf{Structural Constraints} are fairly rigid structures that determine possible micro states of particles that make up the structure,  as in  the case of a cylinder  constraining the motion of a piston, or a skeleton that supports a body.
	
	\item [TD1B]  \textbf{Channeling and Containing constraints} are key forms of contextual causation shaping  microbiological and neural outcomes.  
	\textbf{Channeling constraints} determine where reactants or electrical currents can flow, as in blood capillaries in a body, wires in a computer,  or neural  axons and dendrites in a brain. \textbf{Containing constraints} confine reactants to a limited region, so preventing them from diffusing away and providing the context for reaction networks to function. A key case is  a cell wall. 
	
	\item[TD2A]  
	\textbf{Gating and signalling constraints} Gating constraints control ingress and egress to a container, as in the case of voltage gated ion channels in axons, or ligand gated ion channels in synapses. They function via conformational changes controlled by  voltage differential in the former case, and molecular recognition of ligands in the latter case,  thus underlying cell signalling processes \cite{Berridge}.  
	
	\item [TD2B] \textbf{Feedback control to attain goals} is a cybernetic process where the difference between a goal and the actual state of a system generates an error signal that is fed back to a controller and causes corrective action, as in thermostats and  engine governors \cite{Wiener 1948}. In biology this is \textit{homeostasis},  a crucial feature of physiology at all levels \cite{Guyton 2016}. Because of this closed causal loop, goals determine outcomes. Changing the goals changes both macro  and micro outcomes, as in altering the setting on a thermostat.   In biology, 
	multilevel homeostatic systems are continually responding to internal changes and external perturbations \cite{Billman}. 
	\item[TD3A] \textbf{Creation of New Elements} takes place in two ways.  \textbf{Creation of new lower level elements} occurs in physics when crystal level conditions create  quasiparticles such as phonons that play a key role in dynamics at the electron level \cite{Ellis_2020a_Emerge}. This is what 
	\cite{Gillett} calls a \textit{Downward Constitutive relation}. It occurs in biology when genes are read to create proteins, a contextual process \cite{Gilbert_Epel}  controlled by gene regulatory networks according to higher level needs \cite{Noble 2016}.   \textbf{Creation of new higher level elements}
	  restructures lower level relations and so alters lower level dynamics. In engineering this takes place by manufacturing processes such as making a transistor.  In biology this occurs when cell division takes place at the cellular level, and when an organism gives birth to progeny  at the organism level.  The context of lower level dynamics changes completely in both cases. In the latter case, as Darwin already recognised, sexual selection takes place and determines outcomes, involving very complex social and psychological interactions that alter outcomes at the genetic and physical levels.   
	\item [TD3B] \textbf{Deleting or Altering Lower Level elements} is the complementary process that is crucial in biology. In developmental biology, apoptosis (programmed cell death) plays a key role for example in digit formation (separating fingers and thumbs), while in neural development, synaptic connections are pruned as development takes place \cite{Wolpert}. Cell are specialised to perform specific functions as growth takes place, altering their nature and behaviour.  A fundamental biological process is 
 	\textbf{Adaptive selection due to selection criteria}  which alters either the set of lower level elements by deletion as in Darwinian selection \cite{Campbell_1974} and the functioning of the immune system, or selecting optimal configurations, as in neural network plasticity involved in learning. 
\end{description}
The higher level types of downward causation: \textbf{TD4} (Adaptive selection of goals)
and \textbf{TD5} (Adaptive selection of selection criteria)   
build on these ones \cite{Ellis 2012} \cite{Ellis_2016}. 


The key issue is whether any of these types of downward causation are really causally effective, or just redescriptions in convenient form of microphysical causation. 

\paragraph{Aristotle's kinds of causation} There is an important further point as regards causation.
 As Aristotle pointed out \cite{Bodnar 2018}, there are four kinds of causation that occur in the real world. 
This is discussed by 
(\cite{Juarrero 2002}:2,125-128,143) (\cite{Noble 2016}:176-179)  and  (\cite{Scot 2002}:298-300)
They are
\begin{itemize}
	\item \textbf{Material Cause}:  the physical stuff that is needed for an outcome; the stuff out of which it is made, e.g., the bronze of a statue. In biology this is the physical stuff, the chemical elements as characterised by the periodic table, that make biology possible. 
	
	\item \textbf{Formal Cause}:  which makes anything what it is and no other; 
	the material cause necessary for some outcome must be given the appropriate form through the way in which the material is arranged e.g., the shape of a statue. In biology, this is the structure at each level that underlies function at that level: physiological systems \cite{Guyton 2016} and the underlying biomolecules such as proteins \cite{Petsko_Ringe_2009}. 
	 
	\item \textbf{Efficient Cause}:  The primary source of the change or rest,  the force that brings an action into being; nowadays in the Newtonian case taken to be the effect of forces on inert matter, in the quantum chemistry case, Schr\"{o}dinger's equation.
	
	\item \textbf{Final Cause}: the goal or purpose towards which something aims: ``that for the sake of which a thing is done''. 
\end{itemize}
Physics only considers efficient causes \cite{Juarrero 2002}. Biology however  needs material, formal, and efficient causes. \cite{Hofmeyer 2018 Constructors} gives a careful analysis of how the relation between them can be represented  and how they are are realised in biology,  giving as an example an
enzyme that catalyses a reaction. He explains that  a set of rules, a convention
or code,  forms an interface between formal and efficient
cause. 

All four kinds of causation are needed to determine specific outcomes in social contexts, which is the context within which brains function. Without taking them all into account, one cannot even account for existence of a teapot \cite{Ellis_2005}.

A network of causation is always in action when any specific outcomes occurs. When we refer to `The Cause', we are taking all the others for granted - the existence of the Universe, of laws of physics of a specific nature, and of the Earth for example.

\paragraph{Tinbergen's `Why' questions}  In biology, an alternative view on causation is provided by Tinbergen's four `Why'  questions. \cite{Bateson and Laland 2013} summarise thus:
\begin{quote}\textit{
	 ``Tinbergen pointed out
that four fundamentally different types of problem are
raised in biology, which he listed as `survival value', `ontogeny', `evolution', and `causation'. These problems can be
expressed as four questions about any feature of an organism: What is it for? How did it develop during the lifetime of
the individual? How did it evolve over the history of the
species? And, how does it work?''}
\end{quote}
That is, he raises functional, developmental, evolutionary, and mechanistic issues that all have to be answered in order to give a full explanation of existence, structure, and behaviour of an organism.


\subsection{Multiple Realisability}\label{sec:multiple_realise}
A key point is that multiple realisability plays a fundamental  role in strong emergence \cite{Menzies 2003}. Any particular higher level state can be realised in a multiplicity of ways in terms of lower level states. In engineering or biological cases, a high level need determines the high level function and thus a high level structure  that fulfills it. This higher structure is realised by suitable lower level structures,  but there are billions of ways this can happen.
It does not matter which of the equivalence class of lower level realisations is used to fulfill the higher level need,  as long as it is indeed fulfilled. Consequently you cannot even express the dynamics driving what is happening  in a sensible way at a lower level. \\

Consider for example the statements \textit{The piston is moving because hot gas on one side is driving it } and \textit{A mouse is growing because the cells that make up its body are dividing}. They cannot sensibly be described at any lower level not just because of the billions of lower level particles involved in each case, but because  \textit{there are so many billions of different ways this could happen at the lower level}, this
cannot be expressed sensibly at the proton and electron level. 
The point is the huge number of different combinations of lower level entities can represent a single higher level variable. Any one of the entire equivalence class at the lower level will do. Thus it is not the individual variables at the lower level that are the key to what is going on: it is the equivalence class to which they belong. But that whole equivalence class can be describer by a single variable at the macro level, so that is the real effective variable in the dynamics. This is a kind of interlevel duality:
\begin{equation}\label{eq:Equiva_class}
\{v_\textbf{L} \in \textbf{L} \}\Leftrightarrow \{v_\textbf{i}: v_\textbf{i} \in E_{\textbf{L-1}}(v_\textbf{L-1})\in \textbf{(L-1)}\} 
\end{equation}
where $E_{\textbf{L-1}}(v_\textbf{L-1})$ is the equivalence class of variables $v_\textbf{L-1}$ at Level $\textbf{L-1}$ corresponding to the one variable $v_\textbf{L}$ at Level \textbf{L.} The effective law $\textbf{EF}_\textbf{L}$ at Level \textbf{L} for the variables $v_\textbf{L}$ at that level is equivalent to a law for an entire equivalence class $E_{\textbf{L-1}}(v_\textbf{L-1})$ of variables at Level \textbf{L-1}. It does not translate into an Effective Law for natural variables $v_{\textbf{L-1}}$ \textit{per se} at Level \textbf{L-1}. 

The importance of multiple realisability is discussed in \cite{Menzies 2003} \cite{Ellis_2019_Godel} and \cite{Bishop and Ellis 2020}. 

\begin{quote}
	\textbf{Essentially higher level variables and dynamics}
	\textit{The higher level concepts are indispensible when multiple realisability occurs, firstly because they define the space of data $d_\textbf{L}$ relevant at Level \textbf{L}, and secondly because of  (\ref{eq:Equiva_class}), variables in this space  cannot  be represented  as natural kinds at the lower level. Effective Laws $\textbf{EF}_\textbf{L}$ at level \textbf{L} can only be expressed at level \textbf{L-1} in terms of an entire equivalence class at that level. One can only define that equivalence class by using  concepts defined at level \textbf{L}.}
\end{quote}
To inject reality into this fact, remember that the equivalence class at the lower level is typically characterised by Avagadro's number. 

\subsection{
	 Higher Level Organising Principles}
	 \label{sec:Protectates_HLOP}
A key issue in the discussion is the degree to which higher level dynamics depends on the lower level dynamics. As can be seen from the previous subsections, the nature of biological causation is quite unlike the nature of causation at the underlying physical levels. 
What determines these outcomes then? 

\paragraph{Higher Level Organising Principles} The key idea here is that higher level biological Organising Principles exist that are independent of the underlying lower level dynamics, and shape higher level outcomes. The specific lower level realisation is immaterial, as long as it is in the right equivalence class (Section \ref{sec:multiple_realise}). Generically they form attractors that shape higher level outcomes \cite{Juarrero 2002}152-162; the lower level components come along for the ride, with many biological oscillators being examples (\cite{Noble 2016}:76-86,179).

\paragraph{Protectorates} 
This is parallel to the claim by  \cite{Laughlin_Pines_2000} of existence of classical and quantum protectorates, governed by dynamical rules that characterise emergent systems as such.  They state
\begin{quote}
	``\textit{There are higher organising principles in physics,  such as localization and the principle of continuous symmetry breaking, that cannot be deduced from microscopics even in principle. ... The crystalline state is the simplest known example of a	quantum protectorate, a stable state of matter whose generic low-energy properties are determined by a higher organizing principle and nothing else...  they are transcendent in that they would continue to be true and lead to exact results even if the underlying Theory of Everything was changed.}
\end{quote}
As an example, \cite{Haken} states that profound
 analogies between different systems become apparent
 at the order parameter level, and suggest that the occurrence of order parameters in open systems is a general
law of nature. He characterizes this in terms of a \textit{slaving principle}\footnote{I thank Karl Friston for this comment.} \cite{Haken and Wunderlin}.
 \cite{Green and Batterman 2020} develop this further, citing  the  universality of critical phenomena as a physics case. The Renormalisation Group explanation  extracts structural features that
stabilize macroscopic phenomena irrespective of changes in  microscopic details 

\paragraph{Biology} In biology, such organising principles can be claimed to govern microbiology, physiology, and neuroscience (Sections \ref{sec:nature_biology} and \ref{sec:Plastic_brain}). The idea is that once life exists and evolutionary processes have started, they are what shape outcomes, rather than the underlying physical laws, because they express essential biological needs \cite{Kauffman at home}.

 Physical laws of course \textit{allow} the outcomes to occur: they lie within the \textit{Possibility Space} $\Omega_L$ of outcomes allowed by the physical laws $L$, for instance the proteins enabling all this to occur are characterised by a possibility space of huge dimension, as are the metabolic networks and gene regulatory networks that lead to specific outcomes \cite{Wagner 2014}. But as emergence takes place through developmental processes repeated many many times over evolutionary timescales, it is these principles that determine biological success. Hence \cite{Ginsburg and Jablonka 2019} it is they that determine evolutionary outcomes in an ongoing Evo-Devo process \cite{Carroll 2005} \cite{Carroll 2008}.  They act as attractors for both evolution and for ongoing brain dynamics. \\

 This proposal is supported in multiple ways. 
 
 \textbf{In functional terms,} homeostasis is a central organising principle in all physiology at multiple scales:  ``\textit{It is important to note that homeostatic regulation is not merely the product of a single negative feedback cycle but reflects the complex interaction of multiple feedback systems that can be modified by higher control centers}'' \cite{Billman}. Also  physiological functions acting as dynamical systems have attractors that organise outcomes. For example, this happens in the  
 neural dynamics of cell assemblies (\cite{Scot 2002}:244-248):\begin{quote}
 	``\textit{In Hopfield's formulation, each attractor is viewed as a} pattern \textit{stored non-locally by the net. Each such pattern will have a basin of attraction into which the system can be forced by sensory inputs.''}
 \end{quote}
 Thus cell assemblies form attractors (\cite{Scot 2002}:287). Also Hopfield neural networks converge to attractors in an energy landscape \cite{Churchland  and  Sejnowski}:88-89) and attractor networks are implemented by recurrent collaterals (\cite{Rolls 2016}:75-98).
 
   \textbf{In developmental terms}
 it can be expressed in terms of Waddington's epigenetic landscape \cite{Gilbert 1991} (\cite{Noble 2016}:169,259) which presents much the same idea via cell fate bifurcation diagrams. This is how developmental processes converge on outcomes based in the same higher level organising principles.  
   
   \textbf{In evolutionary terms,} it 
 can be expressed in terms of the adaptive landscape of Sewell Wright \cite{Wright 1932}  \cite{McGhee 2006}, showing how evolution converges to adaptive peaks where these principles are supported to a greater or lesser degree. This viewpoint is supported by much evidence for convergent evolution \cite{McGhee 2011}.  

\paragraph{Neuroscience} There is a huge amount written about neuroscience and biological  psychology, with a vast amount of detail: \cite{Scot 2002} \cite{Purves et al}  \cite{Kandel and Schwartz} \cite{Churchland  and  Sejnowski} \cite{Clark 2016} \cite{Gray} for example. 

The issue is, Can one extract higher level organising principles for the brain from them? 
I believe one can, examples being hierarchical predictive coding \cite{Clark 2013} and the Free Energy Principle \cite{Friston 201O}. I collect them together in the following three sections, looking at how the brain handles the constant influx of new data (Section \ref{sec:brain_open_system}), the issue of constantly adjusting to the environment (Section \ref{sec:Plastic_brain}), and how the brain uses micro level stochasticity to allow macro level agency (Section \ref{sec:stochasticity}). I suggest that it is these principles at the macro level that are the real determinants of what happens, solving the puzzle of how ordered outcomes
can emerge in the context of an open system, where the microdynamic states of an individual brain cannot in principle determine future outcomes because they do not have the data necessary to do so. 

If that is correct,  these principles reach down to determine micro level outcomes via the various mechanisms outlined in Section \ref{sec:causation_types}. 
 Furthermore they are themselves attractors in evolutionary space: they will tend to come into existence because they enhance prospects of reproductive success \cite{Ginsburg and Jablonka 2019}.


\section{The Predictive Brain: Brains as open systems}\label{sec:brain_open_system}

 Each human body, and each brain, is an open system.  This is where the difference between synchronic and diachronic emergence is crucial. It has two aspects: our brains are not made of the same matter as time progresses (Section \ref{sec:matter_change}), and new information is coming in all the time and altering our brain states (Section \ref{sec:brain_predictive_mechanisms}).   The way this is interpreted depends on the fact that our brain is an emotional brain (Section \ref{sec:emotional brain}) and a social brain (Section \ref{sec:social brain}). Language and symbolism enables abstract and social variables to affect outcomes (Section \ref{sec:symbolic brain}). 
 Consequently the microphysical state of a specific person's brain is unable as a matter of principle to predict their future brain states (Section \ref{sec:cannot_predict}) Predictive brains that can handle this situation are attractor states for brain evolutionary development.
 
\subsection{
	Matter and Metabolism: We are not the same molecules}\label{sec:matter_change}
Because we are open systems \cite{Peacock 1989}, the human body at time $t_2 > t_1$ is not made of the same material particles as it was at time $t_1$. Thus what happens in life is like the case of a candle (\cite{Scot 2002}:303):
\begin{quote}
	``\textit{As a simple example of an open system, consider the flame of a candle. .. Because the flame is an open system, a relation $P_1 \rightarrow P_2$ cannot be written - even ``in principle''- for the physical substrate. This follows from the fact that the physical substrate is} continually changing. \textit{The molecules of air and wax vapour comprising the flame at time $t_2$ are entirely different from those at time $t_1$. Thus the detailed speeds and positions of the molecules present at time $t_2$ are unrelated to those at time $t_1$. What remains constant is the flame itself - a} process.''
\end{quote}
\textbf{Body maintenance}: A balance between protein synthesis and protein degradation is required for good health and normal protein metabolism. 
\href{https://en.wikipedia.org/wiki/Protein_turnover}{Protein turnover}  is the replacement of older proteins as they are broken down within the cell, so the atoms and elementary particles making up the cell change too.  Over time,  the human body is not even made up of the same particles: they turn over completely on a timescale of 7 years \cite{EDen et al 2011} \cite{Toyama  and Hetzer 2013}

\paragraph{The brain} Neuroscientist Terence 
Sejnowski states:\footnote{\href{https://www.edge.org/response-detail/10451}{https://www.edge.org/response-detail/10451}} \begin{quote}
	`\textit{`Patterns of neural activity can indeed modify a lot of molecular machinery inside a neuron. I have been puzzled by my ability to remember my childhood, despite the fact that most of the molecules in my body today are not the same ones I had as a child; in particular, the molecules that make up my brain are constantly turning over, being replaced with newly minted molecules. ''}

\end{quote}
 Metabolic networks ensure the needed replacements take place on a continuous basis, despite stochasticity at the molecular level (Section \ref{sec:stochasticity}). This is where multiple realisability plays a key role (Section \ref{sec:multiple_realise}). 

\paragraph{Conclusion} \textbf{\textit{Initial data for the specific set of particles making up a specific brain at time $t_1$ cannot determine emergent outcomes uniquely for that brain over time, for it is not made of the same set of particles at time $t_2\gg t_1$.}}  

\subsection{Dealing with New Information: The Predictive Brain}\label{sec:brain_predictive_mechanisms}
That effect of course takes time. The very significant immediate ongoing  effect of being an open system is that incoming sensory information conveys masses of new data on an ongoing basis. This new data may contain surprises, for example a ball smashes a window. The brain has to have  mechanisms to deal with such unpredictability: the previously stored data at the microphysics level cannot do so, as it does not take this event into account. 

\paragraph{Hierarchical predictive coding} 
Indeed, the brain has developed mechanisms to make sense of the unpredictable inflow of data and best way react to it \cite{Clark 2013} \cite{Clark 2016} \cite{Hohwy 2013} \cite{Hohwy 2016} \cite{Szafron 2019}. 
Andy Clark explains \cite{Clark 2013}:
\begin{quote}
	``\textit{Brains, it has recently been argued, are essentially prediction machines. They are bundles of cells that support perception and
		action by constantly attempting to match incoming sensory inputs with top-down expectations or predictions. This is achieved using a
		hierarchical generative model that aims to minimize prediction error within a bidirectional cascade of cortical processing. Such
		accounts offer a unifying model of perception and action, illuminate the functional role of attention, and may neatly capture the
		special contribution of cortical processing to adaptive success. This `hierarchical prediction
		machine' approach offers the best clue yet to the shape of a unified science of mind and action.''}
\end{quote} 
In brief, following up Ross Ashby's notion that  ``\textit{the whole function of the brain is summed up in error	correction,}'' the following takes place in an ongoing cycle:
\begin{description}
	\item[PB1] \textbf{Hierarchical generative model} \textit{The cortex uses a hierarchical model to  generate predictions of internal and external conditions at time $t_2$ on the basis of data available at time $t_1$.}
	
	\item[PB2] \textbf{Prediction error and attention} \textit{During the interval $[t_1,t_2]$ sensory systems (vision, hearing, somatosensory) receive new  information on external conditions and internal states
	 At time $t_2$, nuclei in the thalamus compare the predictions with the incoming data. If it exceeds a threshold, an error signal (`surprisal') is sent to the cortex to   update its model of the internal and external situation (Bayesian updating), and focus attention on the discrepancy.} 
	 
	\item[PB3] \textbf{Action and outcomes} \textit{The updated model is used to plan and implement action. The impact of that action on the external world provides new data that can be used to further update the model of the external world (active intervention).   }  
\end{description}
This is an interlevel information exchange as described by (\cite{Rao and Ballard 1999}:80):
\begin{quote}
	``\textit{Prediction and error-correction cycles occur concurrently throughout the
		hierarchy, so top-down information influences lower-level estimates, and bottomup
		information influences higher-level estimates of the input signal''}.
\end{quote}
The outcome \cite{Hohwy 2007} is (as quoted in \cite{Clark 2013}), 
\begin{quote}
	``\textit{The generative model
providing the ``top-down''predictions is here doing much of
the more traditionally ``perceptual'' work, with the bottom up driving signals really providing a kind of ongoing feedback
on their activity (by fitting, or failing to fit, the
cascade of downward-flowing predictions). This procedure
combines``top-down'' and ``bottom-up'' influences in an
especially delicate and potent fashion, and it leads to the
development of neurons that exhibit a  ``selectivity that is
not intrinsic to the area but depends on interactions
across levels of a processing hierarchy''} (\cite{Friston 2003}, p.1349). \textit{Hierarchical predictive coding delivers, that is
to say, a processing regime in which context-sensitivity is
fundamental and pervasive}''.
\end{quote}
\paragraph{Perception} Consequently, perception is a predictive affair \cite{Purves}. Helmholz's inverse problem (how to uniquely determine a 3-d world from a 2-d projection) is solved by filling in missing information on the basis of our expectations.  (\cite{Kandel 2012}:202-204) gives a overview of how this understanding originated with Helmholz, who called this top-down process of hypothesis testing \textit{unconscious inference}, and was developed by  Gombrich in his book \textit{Art and Illusion} \cite{Gombrich}. \cite{Kandel 2012} (pages 304-321)  emphasizes the top-down aspect of this process, and its relation to memory. \cite{Purves}(pp.120-124) describes how he came to the same understanding (see also page 221).

\paragraph{Action} The relation to action is given by \cite{Friston 2003} \cite{Friston et al 2009} \cite{Clark 2013}. It is described thus by (\cite{Hawkins}:158)
\begin{quote}
	``\textit{As strange as it sounds, when your own behaviour is involved, your predictions
		not only precede sensation, they determine sensation. Thinking of going to the
		next pattern in a sequence causes a cascading prediction of what you should
		experience next. As the cascading prediction unfolds, it generates the motor
		commands necessary to fulfill the prediction. Thinking, predicting, and doing are
		all part of the same unfolding of sequences moving down the cortical hierarchy.''}
\end{quote}
\cite{Seth 2013} summarised the whole interaction thus: 
\begin{quote}
	``\textit{The concept of Predictive Coding (PC) overturns classical notions of perception
		as a largely `bottom-up' process of evidence accumulation
		or feature detection, proposing instead that perceptual
		content is specified by top-down predictive signals that
		emerge from hierarchically organized generative models of
		the causes of sensory signals. According to PC, the brain is
		continuously attempting to minimize the discrepancy or
		`prediction error' between its inputs and its emerging
		models of the causes of these inputs via neural computations
		approximating Bayesian inference. Prediction errors can be minimized
		either by updating generative models (perceptual inference
		and learning; changing the model to fit the world) or
		by performing actions to bring about sensory states in line
		with predictions (active inference; changing the world to
		fit the model''}
\end{quote}
This is a very brief sketch of a very complex program,  summarised in Andy Clark's book \textit{Surfing Uncertainty} \cite{Clark 2016} and in \cite{Miller and Clark 2018}. 
Nothing here contradicts the mechanisms discussed in depth in texts such as \cite{Purves et al} \cite{Kandel and Schwartz} \cite{Churchland  and  Sejnowski}. Those texts set the foundations for what is proposed above, but do not develop these aspects in depth. For example \cite{Kandel and Schwartz} has just one relevant section: ``Visual perception is a creative process'' (page 492).

Thus the viewpoint put here accepts the mechanisms discussed in those books (and the underlying physics), and puts them in a larger context that emphasizes overall organising features that are crucial in enabling the brain to function in the face of uncertainty.  

However there are  three further important aspects to be taken into account. 

\subsection{The emotional brain}\label{sec:emotional brain}
A first further crucial aspect of our brains is that they are \textbf{emotional brains}. The understandings and actions enabled by the predictive mechanisms mentioned above are crucially affected and shaped by affective (emotional) states. 

The cognitive science paradigm of purely rational choice is not the way the real brain works. Emotion has key effects on cognition \cite{Damasio} and behaviour  \cite{Panksepp} \cite{Purves et al} \cite{Panksepp and Biven} \cite{Gray}. 
\begin{description}
	\item[EB1] \textbf{The emotional brain} \textit{Both primary (innate) and secondary (social) emotions play a key role in guiding cognition and focusing attention.}
\end{description}
The predictive coding paradigm can be extended (\cite{Clark 2016}:231-237) to include this case.  \cite{Seth 2013} says the following
\begin{quote}
	``\textit{The concept of the brain as a prediction machine has
		enjoyed a resurgence in the context of the Bayesian brain
		and predictive coding approaches within cognitive science.
		To date, this perspective has been applied primarily
		to exteroceptive perception (e.g., vision, audition), and
		action. Here, I describe a predictive, inferential perspective
		on interoception: `interoceptive inference' conceives
		of subjective feeling states (emotions) as arising from
		actively-inferred generative (predictive) models of the
		causes of interoceptive afferents. The model generalizes
		`appraisal' theories that view emotions as emerging
		from cognitive evaluations of physiological changes ... interoceptive inference involves hierarchically
		cascading top-down interoceptive predictions that counterflow
		with bottom-up interoceptive prediction errors. Subjective
		feeling states - experienced emotions - are
		hypothesized to depend on the integrated content of these
		predictive representations across multiple levels ''}
\end{quote}

\noindent \cite{Miller and Clark 2018} develop this crucial emotional relationship to cortical activity in depth, using the predictive coding framework:
\begin{quote}
	``\textit{But
		how, if at all, do emotions and sub-cortical contributions fit into this emerging picture?
		The fit, we shall argue, is both profound and potentially transformative. In the picture
		we develop, online cognitive function cannot be assigned to either the cortical or
		the sub-cortical component, but instead emerges from their tight co-ordination. This
		tight co-ordination involves processes of continuous reciprocal causation that weave
		together bodily information and `top-down' predictions, generating a unified sense of
		what's out there and why it matters. The upshot is a more truly `embodied' vision of
		the predictive brain in action.''}
\end{quote} 
As well as influencing immediate functioning of the brain, affect relates crucially to brain plasticity and so to changes in brain micro structure  (Section \ref{sec:Neural_Darwin}).
\subsection{The social brain}\label{sec:social brain}
Second, a crucial aspect of our brains is that they are \textbf{social brains}: we are evolved to live in a social context, which has key influences on our lives and minds as   the brain receives data and responds to the situation around.  Sociality appears to be a main driver for human brain evolution   \cite{Dunbar 1998} \cite{Dunbar 2003} and results in social cognition (\cite{Purves et al}:359-392) and cognitive neuroscience \cite{Cacciopo}. This again crucially affects how we handle the incoming information.

\paragraph{The advantage of social brains} Living in cooperative groups greatly enhanced our ancestors survival prospects \cite{Harari 2014} enabling the rise of cooperative farming, culture, and technology, which then was the key to the emergence of civilisation that enabled our dominance over the planet \cite{Bronowski 2011}. A social brain was needed for social cohesion to emerge: the cognitive demands of living in complexly bonded social groups selected increasing executive brain (neocortical) size  \cite{Dunbar 1998a} \cite{Dunbar 2014}.

\paragraph{The nature of the social brain: Theory of Mind}
It is not just a matter of being cooperative and able to communicate: central to the social brain is the ability known as ``theory of mind'' (ToM) \cite{Dunbar 1998a}. 
It is very important that we can read other peoples minds (understanding their intentions) - which we do on an ongoing basis \cite{Frith 2013}. We all have a theory of mind \cite{Frith and Frith 2005}. Its cortical basis is discussed by  \cite{Frith 2007}, but additionally   
it has a key precortical base related to the primary emotional systems identified by \cite{Panksepp}, namely the very strong  emotional need to belong to a group \cite{Panksepp and Biven} \cite{Ellis and Toronchuk 2013}  
 \cite{Stevens and  Price 2015} \\
 
Its evolutionary basis is discussed by \cite{Donald 1991} \cite{Tomasello 2009} \cite{Dunbar 2014}. It is summed up by (\cite{Donald 2001}:86-87) as follows:
\begin{quote}
	``\textit{Our normal focus is social, and social  awareness is highly conscious, that is, it heavily engages our conscious capacity... Conscious updating is vital to social life ... One might even make the case that consciousness- especially our lightning fast, up-to-date, socially attuned human consciousness - is the evolutionary requirement for both constructing and navigating human culture. It remains the basis, the sine qua non, for all complex human interactions}''.
\end{quote}
Michael Tomasello agrees, as is evident in the title of his book \textit{The Cultural Origin of Human Cognition} \cite{Tomasello 2009}.


\paragraph{Relation to predictive coding}
The description of the social brain in terms of the predictive processing paradigm is presented by  \cite{Constant et al} through the concept of the extended mind:\footnote{See also \cite{Kirchhoff} and \cite{Hesp}.}
\begin{quote}
	``\textit{Cognitive niche construction is construed as a form of instrumental intelligence, whereby organisms create and
		maintain cause-effect models of their niche as guides for fitness influencing behavior. Extended mind theory
		claims that cognitive processes extend beyond the brain to include predictable states of the world  that function
		as cognitive extensions to support the performance of certain cognitive tasks. Predictive processing in cognitive
		science assumes that organisms (and their brains) embody predictive models of the world that are leveraged to
		guide adaptive behavior. On that view, standard cognitive functions - such as action, perception and learning -
		are geared towards the optimization of the organism's predictive (i.e., generative) models of the world. Recent
		developments in predictive processing - known as active inference - suggest that niche construction is an
		emergent strategy for optimizing generative models. 
	}
\end{quote}
Those models include models of social context and of other minds, characterised via cultural affordances \cite{Ramstaed et al culture}. \cite{Veissiere} state
\begin{quote}
	``\textit{We argue that human
		agents learn the shared habits, norms, and expectations of their culture through immersive
		participation in patterned cultural practices that selectively pattern attention and behaviour.
		We call this process``Thinking Through Other Mind'' (TTOM) - in effect, the process of
		inferring other agents' expectations about the world and how to behave in social context. }''
\end{quote}
Then downward causation from the social environment changes the brain:
\begin{quote}
	``\textit{The brain only has direct access to the way its sensory states fluctuate (i.e., sensory input),
	 and not the causes of those inputs, which it must learn to guide adaptive action  -
	 where `adaptive' action solicits familiar, unsurprising (interoceptive and exteroceptive)
	 sensations from the world. The brain overcomes this problematic seclusion by matching the
	 statistical organization of its states to the statistical structure of causal regularities in the
	 world. To do so, the brain needs to re-shape itself, self-organizing so as to expect, and be
	 ready to respond with effective action to patterned changes in its sensory states that
	 correspond to adaptively relevant changes `out there' in the world''}
\end{quote}
The sociology of this all is discussed  by \cite{Berger} and \cite{Berger and Luckmann 1991}. Overall, one can summarise as follows: 

\begin{description}
	\item[SB1] \textbf{The social brain} \textit{Because we live in a social world we are very socially aware. We have a social brain which shapes our responses to incoming data in crucial ways on the basis of social understandings, which are continually changing over time}. 
\end{description}
Theory of mind is based in prediction, and is a routine part of everyday life \cite{Frith 2013}.

\subsection{The symbolic brain}\label{sec:symbolic brain}
Third, a key feature of the social brain is its ability to engage in spoken and written language, and more generally to engage in symbolism. This adds in a whole new category of incoming information that the brain has to take into account and respond to.

\paragraph{Language} 
 A key step in evolution of mind is developing language.     \cite{Dunbar 1998a} suggests its prime function is to enable exchange of information regarding bonding in the social group. It is a product of a mind-culture symbiosis 
  (\cite{Donald 2001}:11,202) and forms the basis of culture (\cite{Donald 2001}:274), symbolic technologies (\cite{Donald 2001}:305), as well as cultural learning  (\cite{Tomasello 2009}:6) and inheritance (\cite{Tomasello 2009}:13). 
   \cite{Ginsburg and Jablonka 2019}
Language enables sharing ideas and information over time and distance, and enables the social and psychological power of stories \cite{Gottschall 2012}. 
 
  \paragraph{Abstract and social variables}
In evolutionary terms, the transition to the symbolic species \cite{Deacon 1997} enabled abstract causation  \cite{Ellis_Kopel_2019} to occur, 
which \textit{inter alia} involves social interactions and abstract concepts such as the amount of money in my bank account and the concept of a closed corporation \cite{Harari 2014}. Thus not all the relevant variables are physical variables; some are abstract variables resulting from social interactions \cite{Berger} \cite{Berger and Luckmann 1991} which  are causally effective. 

\paragraph{Higher order predictability}
Symbolism and abstract reasoning greatly increases our power of prediction: we can simulate situations offline, rather than having to enact them to see what the consequences are. It also greatly increases the complexity of our responses to incoming social data, which are interpreted in the light of the social context 
\cite{Berger} \cite{Berger and Luckmann 1991} \cite{Donald 2001} \cite{Frith 2013}

\begin{description}
	\item[SB2] \textbf{The symbolic brain} \textit{Human social interaction is based in language, in turn based in our symbolic ability. This ability transforms the way our minds interpret much incoming data, as well as allowing internal cognitive processes that are a major causal factor in our individual and social lives.}
\end{description}
This is the fundamental mechanism by which the brain operates at a macro level, for which there is much evidence. Again one can claim that this is the way the brain operates as a matter of fact,  it is not just the way we think it operates. Causation at this level is real: the whole of society depends on it. 
  
This will play an important role in Section \ref{sec:whole_universe} because it relates to the interaction of the brain to the outside world. 
 
\subsection{The Dynamics of the Open Brain}\label{sec:open_brain_dynamics}
An individual brain considered as an entity on its own is an open system, and has been adapted to handle the problems this represents in successfully navigating the world. This rather than initial brain micro data determines it specific outcomes as time progresses.

\paragraph{Microphysics data for brain states}\label{sec:cannot_predict}
Consider a specific individual brain at time $t_1$. During a time interval $[t_1,t_2]$, the initial brain microphysics data ${\cal D}(t_1)$   is added to by new data ${\cal D}_{ext}(t)$ coming from the environment after $t_1$. The data ${\cal D}(t_2)$ at a later time $t_2>t_1$ is not predictable even in principle from ${\cal D}(t_1)$. Hence the microphysics evolution is undetermined by data   ${\cal D}(t_1)$, even in principle. You may for example see a car crash at a time $t_3>t_1$ that alters all the future brain states; but your brain did not know that was going to happen.  

Thus the brain as an open system receives unexpected information and handles it in a predictive way. The initial state of the brain obviously cannot determine these outcomes as it has no control over what the incoming data will be.  This is the key outcome of the difference between synchronic and diachronic emergence. 

\begin{quote}
	\textbf{The brain is an open system \textit{Initial micro data of a brain state at one moment cannot possibly determine what it will do at a later time, 
			not just because new matter comes in and replaces old, but also because  new information comes in from outside and alters outcomes. The initial data at time $t_1$ cannot know what the initial data at time $t_2$ will be and hence cannot determine specific later brain outcomes. The brain handles this uncertainty via the predictive brain mechanisms \textbf{PB1-PB3}, \textbf{EB1}, \textbf{SB1-SB2} outlined above.}}
\end{quote}
The physicalist gambit is to say ah yes, but microphysics determines uniquely the evolution of all the other systems the brain is interacting, so the system as a whole is determined by the microphysics dynamics alone. I respond to that proposal in Section \ref{sec:whole_universe}.

\paragraph{Predictive Brain Mechanisms as Attractor states}\label{sec:predictive_mechanism_attractors}

Evolutionary processes will hone in on these predictive brain mechanisms as attractor states. This occurs via the mechanism of exploration and selective stabilisation  recognised independently by Changeaux and by Edelman (\cite{Ginsburg and Jablonka 2019}:119-123,247-248).

Thus these mechanisms can be claimed to be Higher Order Principles (see Section \ref{sec:Protectates_HLOP}) for brain structure and function. It is their remarkable properties that shape brain structure, and its functioning in the
face of the unpredictable flow of incoming data. 

\section{The Learning Brain: Plasticity  and Adaptation}\label{sec:Plastic_brain}
 In carrying out these responses to incoming information, remembering and learning takes place; indeed this is a pre-requisite for functioning of predictive brain mechanisms. This adds a new dimension to the effects  just discussed: not only is the new data unpredictable, but also brain structure is changed in ways affected by that inflow of new data. Thus the context for microphysics outcomes - the specific set of constraints determining electron and ion flow possibilities -  is also different at the later time.
 
 Plasticity at the macro level  as the brain adapts to its environment, remembers, and learns \cite{Gray}   is enabled by corresponding changes at the micro level  as neural networks weights change \cite{Kandel and Schwartz} \cite{Churchland  and  Sejnowski}. Thus changes  take place at  the micro level (Section \ref{sec:plastic_micro}) driven by incoming data at the macro level, and resulting in plasticity at the macro level (Section \ref{sec:plastic_macro}).  Because brain neural nets are changing all the time, the context for outcomes of the underlying physics is also changing all the time (Section \ref{sec:brain_change}) and is not predictable from the initial brain physical microstate.
\subsection{Plasticity at the micro level}\label{sec:plastic_micro}
Learning takes place by change of connectivity and weights in neural networks at the neuronal level  \cite{Kandel and Schwartz} \cite{Churchland  and  Sejnowski}, taking place via gene regulation at the cellular level \cite{Kandel_memory}. This alters the context within which electron and ion flows take place in neural networks and in particular at synapses, thereby shaping outcomes of the underlying universal physical laws.

\paragraph{Developmental processes}
This plasticity occurs particularly when brain development is taking place. Random initial connections are refined (\cite{Wolpert}: \S11) and new experiences can modify the original set of neuronal connections (\cite{Gilbert 1991}:642) while the brain is responding to the surrounding environment (\cite{Purves}:\S2-\S5, 229).

\paragraph{Learning Processes} Erik Kandel explored the mechanism of  learning in depth. He identified gene regulatory process related to learning  \cite{Kandel_memory}\textit{\begin{quote}
		 ``Serotonin acts on specific
receptors in the presynaptic terminals of
the sensory neuron to enhance transmitter
release. ...  during long-term memory storage, a
tightly controlled cascade of gene activation is
switched on, with memory-suppressor genes
providing a threshold or checkpoint for memory
storage ... With both implicit and explicit memory
there are stages in memory that are encoded
as changes in synaptic strength and that
correlate with the behavioral phases of
short- and long-term memory''
\end{quote}}
\paragraph{The relation to physics}
These changes alter the context within which the underlying physics operates. Changing constraints at the microphysics level 
 is the mechanism of downward causation to that level \cite{Ellis_Kopel_2019}. This determines what dynamics actually takes place at the ion/electron level, which of course the fundamental laws by themselves cannot do. The outcomes are determined by biological context in this way.  

\begin{description}
	\item[LB1] \textbf{The Developing and Learning Brain} \textit{The brain is plastic at the micro level, as  development and learning takes place. Neural network connections and weights are altered via  gene regulatory processes.}
\end{description}
Thus neural network learning \cite{Churchland  and  Sejnowski} - a real causal process at each network level - alters electron outcomes and so later psychological level dynamics. 


\subsection{Plasticity at the  macro level}\label{sec:plastic_macro}
Eric Kandel states ``\textit{One of the most remarkable aspects of an animal's behavior is the ability to modify
that behavior by learning}'' \cite{Kandel_memory}, and emphasizes that 
social factors affect this learning. \cite{Kandel psych} gives five principles for psychotherapy that make this clear. For those who are skeptical of psychotherapy,  replace that word with `teaching' or `coaching' in the following, and its crucial meaning still comes through. 
 
\begin{description}
 	\item [Kandel Principle 1] All mental processes, even the most complex
 	psychological processes, derive from operations of
 	the brain. The central tenet of this view is that what we 	commonly call mind is a range of functions carried out
 	by the brain. The actions of the brain underlie not only
 	relatively simple motor behaviors, such as walking and
 	eating, but all of the complex cognitive actions, conscious
 	and unconscious, that we associate with specifically
 	human behavior, such as thinking, speaking, and
 	creating works of literature, music, and art.
 	
 	\item [Kandel Principle 2]  Genes and their protein products are important
 	determinants of the pattern of interconnections
 	between neurons in the brain and the details of their
 	functioning. Genes, and specifically combinations of
 	genes, therefore exert a significant control over behavior. ... the transcriptional function of a gene - the ability of a
 	given gene to direct the manufacture of specific proteins
 	in any given cell - is, in fact, highly regulated, and this
 	regulation is responsive to environmental factors ... the regulation of gene expression by
 	social factors makes all bodily functions, including all
 	functions of the brain, susceptible to social influences.
 	
 		\item [Kandel Principle 3] Behavior itself can also modify gene expression. Altered genes do not, by themselves, explain
 		all of the variance of a given major mental illness.
 		Social or developmental factors also contribute very importantly.
 		Just as combinations of genes contribute to
 		behavior, including social behavior, so can behavior
 		and social factors exert actions on the brain by feeding
 		back upon it to modify the expression of genes and thus
 		the function of nerve cells. Learning  ...
 		 produces alterations
 		in gene expression. 
 	
 		\item [Kandel Principles 4/5]
 How does altered gene expression lead to the stable
 alterations of a mental process? 
  Alterations in gene expression induced by
 		learning give rise to changes in patterns of neuronal
 		connections. These changes not only contribute to the
 		biological basis of individuality but 
  strengthen the effectiveness
 of existing patterns of connections,  also
 changing cortical connections to accommodate new patterns
 of actions....		
 resulting in long-lasting
 effect on the 	the anatomical pattern of interconnections between
 nerve cells of the brain.
\end{description}
\paragraph{The hierarchical predictive coding view}
The way this all fits into the predictive coding viewpoint discussed in the last section is explained by \cite{Rao and Ballard 1999}. \\

Overall the outcomes can be summarised thus:
\begin{description}
	\item[LB2] \textbf{The Learning Brain} \textit{The brain is plastic at the macro level as learning takes place, supported by plasticity at the micro level. Learning at the macrolevel responds to social and psychological variables. 
	}
\end{description}

\subsection{The ever adapting  brain}\label{sec:brain_change}
The previous section emphasized, in the case of a single brain, that because of incoming data, the microstate at time $t_2$ cannot be predicted from initial data at time $t_1<t_2$ because it does not include this incoming data. This section emphasizes that in addition, the micro level constraints are changed because neural network wiring or weights will have changed as the brain adapts at both macro and micro levels to ongoing environmental events and changes. So not only is the data different than expected because the brain is an open system, but the dynamical context for the underlying physics is different too.

\begin{quote}
	\textbf{The brain is an adaptive system.  \textit{Individual brain structure changes in response to incoming data. As new information comes in, neural network weights are continually changed via gene regulation. This change of context alters constraints  in the underlying Lagrangian, and so changes the context for future physical interactions. None of this can be determined by the initial brain micro data at time $t_1$, as these changes are shaped by data that has come in since then.}}

\end{quote}
This is a further reason why diachronic emergence is crucially  different from synchronic. 

\paragraph{Adapting and learning  brains as attractor states}
Evolutionary processes will hone also in on these learning brain mechanisms as attractor states, via the mechanism of exploration and selective stabilisation  recognised independently by Changeaux and by Edelman (\cite{Ginsburg and Jablonka 2019}:119-123,247-248).



\section{The Stochastic Brain and Agency}\label{sec:stochasticity}
A key feature undermining physicalist determinism of brain states is the stochasticity that occurs in biology at the molecular level, which uncouples biology from detailed Laplacian determinism. Section \ref{sec:bio_stochastic} discusses this stochasticity, and Section \ref{sec:stochastic_select} how this opens up the way for selecting desired low level outcomes that will fulfill higher level purposes - one of the key forms of downward causation.  Section \ref{sec:stochastic_brain} discusses how this applies specifically to the brain.   A key way that randomness is used in shaping the brain is Neural Darwinism (Section \ref{sec:Neural_Darwin}).
The issue of how agency is possible arises, and this is essentially via multi level causal closure that takes advantage of this selective process (Section \ref{sec:agency_causal_closure}). 

This shows how biological stochasticity opens up the way to higher level biological needs acting as attractors that shape brain dynamics, rather than brain outcomes being the result purely of deterministic or statistical lower level physical dynamics. All of this can again be traced out at the underlying physics level, but it is the biology that is the essential causal factor through setting the context for physical outcomes.

\subsection{Biology and stochasticity}\label{sec:bio_stochastic}
There is massive stochasticity at the molecular level in biology. This undoes Laplacian determinism at the micro level: it decouples molecular outcomes from details of initial data at the molecular level. How then does order emerge? By biology harnessing this stochasticity to produce desirable higher level results, as happens for example in the case of molecular machines \cite{Hoffman} and the adaptive immune system \cite{Noble and Noble 2018 Stochasticity}. 

\paragraph{Stochasticity and molecular machines} As described in Hoffman's book  \textit{Life's ratchet: how molecular machines extract order from chaos} \cite{Hoffman}
 biomolecules live in a cell where a molecular storm occurs. Every molecular machine in our cells is hit by a fast-moving water molecule about every $10^{-13}$ seconds. He states 
  \begin{quote}
	``\textit{At the nanoscale, not only is the molecular storm an overwhelming force, but it is also completely random.}''
\end{quote} 
The details of the initial data molecular positions and momenta) are simply lost. To extract order from this chaos,  ``\textit{one must make a machine that could `harvest' favorable pushes from the random hits it receives.}'' That is how biology works at this level. 

\paragraph{Stochasticity in gene expression} Variation occurs in the expression levels of proteins \cite{Chang et al 2008}. This is a property of the population as a whole not of single cells,  so the distribution curve showing the number of cells displaying various levels of expression is an attractor created by the population (\cite{Noble 2016}:175-176). Promoter architecture
is an ancient mechanism to control expression variation \cite{Sigalova}. Thus one must use stochastic modeling of biochemical and genetic networks (\cite{Ingalls}:280-295) when determining their outcomes.
This affects the detailed physical outcomes of memory processes based in gene regulation \cite{Kandel_memory}. 

\paragraph{Stochasticity in genetic variation} The processes of genetic variation before selection are mutation and recombination \cite{Alberts_etal_07},  drift \cite{Masel 2011},  and migration. They all are subject to stochastic fluctuations. Mutations arise spontaneously at low frequency owing to the chemical instability of purine and pyrimidine bases and to errors during DNA replication \cite{Lodish}. Because evolution is a random
walk in a state space with dimension given by the
number of the different strategies present \cite{Geritz} this shapes evolutionary outcomes in a way that is unpredictable on the basis of microphysics data. 
This has key present day outcomes in terms of ongoing mutations of microbes and viruses on the one hand, and of immune system responses on the other, both of which are based in taking advantage of stochasticity \cite{Noble and Noble 2018 Stochasticity}. The statistics of outcomes however can be studied in terms of evolution over a rugged fitness landscape \cite{Gillespie} \cite{Kauffman} \cite{Felsenstein} \cite{Orr 2005}. Small fluctuations can end up in a different attractor basin  or adaptive peak.

\paragraph{The microbiome} A key factor in physiology is that our bodies contain many billions of microbes that affect bodily functioning and health. This is being studied in depth by the Human Microbiome Project \cite{Peterson} and the Integrative Human Microbiome Project \cite{Integrative 2014} \cite{Integrative 2019}. 

Thousands	of	microbial	species,	possessing	millions	of	genes,	live	within	humans: in the  gastrointestinal and urogenital tracts, in the 
mouth: $10^{18}$, in the nose, in the lungs: $10^9$/ml, on the skin:  $10^{12}$.
This leads to infectious diseases (rheumatic fever, hepatitis, measles, mumps, TB, AIDS, inflammatory bowel disease) and allergic/autoimmune diseases (Asthma, diabetes, multiple sclerosis, Croon's disease).

 Because of the stochasticity in gene 
mutation, recombination, and horizontal gene transfer, and the huge numbers involved, together with the impossibility of setting data to infinite precision (\S\ref{sec:micor_determine}), evolution of specific outcomes is unpredictable in principle, but has major macro level outcomes for individuals.  \ 
\subsection{Stochasticity and selection in biology}\label{sec:stochastic_select}
This level of stochasticity raises a real problem: how could reliable higher levels of biological order, such as functioning of metabolic and gene regulatory networks and consequent reliable development of an embryo \cite{Wolpert}, emerge from this layer of chaos? 

The answer is that evolutionary processes have selected for
biological structures that can successfully extract  order from the chaos. These structures in turn use the same mechanism: they select for biologically desirable outcomes from an ensemble of physical  possibilities presented by this underlying randomness.  Higher level biological needs may be satisfied this way. As stated by \cite{Noble and Noble 2018 Stochasticity}:
\begin{quote}
``\textit{Organisms can harness stochasticity through which they can generate many possible solutions to environmental challenges. They must then employ a comparator to find the solution that fits the challenge. What therefore is unpredictable in prospect can become comprehensible in retrospect. Harnessing stochastic and/or chaotic processes is essential to the ability of organisms to have agency and to make choices.} 
\end{quote}
This is the opposite of the Laplacian dream of  the physical interactions of the underlying particles leading to emergent outcomes  purely on the basis of the nature of those  interactions. It is the detailed structure of molecular machines, together with the lock and key molecular recognition mechanism used in molecular signalling \cite{Berridge}, that enables the logic of biological processes to emerge as effective theories governing dynamics at the molecular level. 
They exist in the form they do because of the higher level organising principles that take over. The emergent levels of order  appear because they are based in higher level organising principles characterising emergent protectorates as described by  \cite{Laughlin_Pines_2000} (see Section \ref{sec:Protectates_HLOP}). 
For example Friston's Free Energy Principle \cite{Friston 201O} \cite{Friston 2012} is such a higher level organising principle. It does not follow from the microphysical laws. In fact all the higher level Effective Theories $\textbf{ET}_\textbf{L}$ (Section \ref{sec:equal_validity_levels}) characterise such Higher Level Organising principles. \\


 \begin{description}
	\item[STB1]\textbf{Variation leads to a variety of states, from which outcomes are selected; }
\textit{ \textbf{States that fulfill biological functions are attractor states for function and hence for evolution and development.}
}\end{description}
This agrees with (\cite{Ginsburg and Jablonka 2019}:245)
\begin{quote}
	``\textit{Biological attractors are usually functional -  the mechanisms enabling them to be reached reliably, in spite of different starting conditions, evolved by natural selection''.}
\end{quote}
This is the process exploration and selective stabilisation mechanism that is described in (\cite{Ginsburg and Jablonka 2019}:119-123,247-248). 
The driving of the process by biological needs is the reason that convergent evolution occurs \cite{McGhee 2011}.


\subsection{The brain and stochasticity}\label{sec:stochastic_brain}

There are various kinds of stochasticity in brain function, apart from the fact that it involves necessarily the stochasticity in molecular dynamics just discussed.

\paragraph{Stochasticity in Neural Activity}
 The neural code is spike chains \cite{Rieke et al} where \cite{Shadlen and Newsoms 1994} the timing of
successive action potentials is highly irregular. Also fluctuations in cortical excitability occur \cite{Stephani 2020}. 
This results in stochasticity in neural outcomes \cite{Glimcher} in contrast to deterministic dynamics, suggesting an organising principle \cite{Stephani 2020}:
\begin{quote}
	``\textit{Brain responses vary considerably from moment to moment, even to identical sensory stimuli. This has been attributed to changes in instantaneous neuronal states determining the system's excitability. Yet the spatio-temporal organization of these dynamics remains poorly understood. 
		.... criticality may represent a parsimonious organizing principle of variability in stimulus-related brain processes on a cortical level, possibly reflecting a delicate equilibrium between robustness and flexibility of neural responses to external stimuli.''}
\end{quote}
This stochasticity allows higher level organising principles such as attractors to shape neural outcomes in decision making in the brain \cite{Rolls and Deco 2010} \cite{Deco  2009}. The higher level structure of 
 attractor networks \cite{Rolls 2016}:95-134) determines outcomes. 
A particular case where a randomisation and selection process is used is Boltzmann machines and annealing (\cite{Churchland  and  Sejnowski}:89-91). This demonstrates the principle that 
\textit{stochasticity greatly enhances efficiency in reaching attractor states} \cite{Palmer 2020}. 

\paragraph{Creativity} A key feature of mental life is creativity, which has transformed human life both through inventiveness in science (Maxwell, Turing, Bardeen, Townes, Cormack, and so on) and in commerce (Gates, Jobs, Zuckerberg, Bezos, and so on). It has been   proposed (\cite{Rolls 2016}:137) that the possibility of creativity is an outcome of stochasticity due to random spiking of neurons,  resulting in a brain state being able to switch from one basin of attraction to another. 

\paragraph{The gut-brain axis} 
The body microbiome (Section \ref{sec:bio_stochastic}) has a key influence on the brain. Effects are as follows  \cite{Cryan}  
\begin{quote}
	``\textit{The microbiota and the brain communicate with each other via various routes including the immune system, tryptophan metabolism, the vagus nerve and the enteric nervous system, involving microbial metabolites such as short-chain fatty acids, branched chain amino acids, and peptidoglycans. Many factors can influence microbiota composition in early life, including infection, mode of birth delivery, use of antibiotic medications, the nature of nutritional provision, environmental factors, and host genetics. At the other extreme of life, microbial diversity diminishes with aging. Stress, in particular, can significantly impact the microbiota-gut-brain axis at all stages of life. Much recent work has implicated the gut microbiota in many conditions including autism, anxiety, obesity, schizophrenia, Parkinson's disease, and Alzheimer's disease.''}
\end{quote}
It is also involved in neurodegenerative disease \cite{Rosario}. Because of the unpredictability of how the microbiome will develop, both due to the stochasticity of its genetic mutation and the randomness of the microbes imported from the environment, the specific outcomes of these interactions are unpredictable from initial micro biological data in an individual body, and hence \textit{a fortiori} from knowledge of the details of the underlying physical level.  One can however study the statistics of molecular evolution over the mutational landscape  \cite{Gillespie} \cite{Kauffman}. 

Note that not all the key  factors determining outcomes are purely microbiological or physiological: stress, a mental state, is a key factor in its dynamics.

\subsection{Neural Plasticity and Neural  Darwinism}\label{sec:Neural_Darwin}
As well as changing neural network weights \cite{Churchland  and  Sejnowski} via gene regulatory networks\cite{Kandel_memory}, neural plasticity during development involves  pruning connections that were initially made randomly \cite{Wolpert} as learning takes place. \\

An  important way variation and selection happens is via
\textbf{Neural Darwinism} (or Neuronal Group Selection) 
\cite{Edelman 1987} \cite{Edelman 1993}. This is a process where neural connections are altered by neuromodulators such as dopamine and serotonin that are diffusely spread from precortical nuclei to cortical areas via `ascending systems'. They then modify weights of all neurons that are active at that time, thus at one shot strengthening or weakening an entire pattern of activation - a vary powerful mechanism.  This mechanism  (\cite{Ginsburg and Jablonka 2019}:119-123,247-248) was also discovered by \cite{Changeux and Danchin 1976}.
\cite{Seth and Baars} describe these processes thus:
\begin{quote}
	``\textit{In the brain, selectionism applies both to neural development and to moment-to-moment functioning.
	Edelman postulates two overlapping phases of developmental and experiential variation
	and selection. The first is the formation during development of a primary repertoire of many neuronal
	groups by cell division, migration, selective cell death, and the growth of axons and dendrites.
	This primary repertoire of neurons is epigenetically constructed through a suite of genetic and environmental influences, and generates a high level of diversity in the nascent nervous
	system.
	The second, experiential, phase involves the dynamic formation from this primary repertoire of
	a secondary repertoire of functional neuronal groups, by the strengthening and weakening of synapses
	through experience and behavior. This phase involves the selective amplification of functional
	connectivities among the neurons produced in the first phase, with which it overlaps. In
	this manner, an enormous diversity of anatomical and functional circuits is produced}''
\end{quote}
This provides a key mechanism for experientially based  selection of connectivity patterns.

\paragraph{Primary Emotions}
An important feature of Edelman's theory is that the subcortical nuclei involved, as well as the neuromodulators, are precisely the same as are involved in Jaak Panksepp's primary emotional systems \cite{Panksepp} (Section \ref{sec:emotional brain}). Hence the theory is in fact a theory of \textit{\textbf{Affective Neural Darwinism}} \cite{Ellis and Toronchuk 2005}
\cite{Ellis and Toronchuk 2013}, making clear the importance of affect (emotion) for  brain plasticity and learning. 

\begin{description}
	\item[STB2] \textbf{Stochasticity and Neural Darwinism} \textbf{\textit{Brain plasticity is affected by neuromodulators diffusely projected to the cortex from nuclei in subcortical arousal system via ascending systems, selecting neuronal groups for strengthening or weakening. In this way emotions affect neural plasticity. }}
\end{description}
This is a key way that interactions at the social level
reach down to alter brain connections and hence the  context of physical interactions in the brain. It is a specific example of the `vary and select' topdown process (Section \ref{sec:causation_types}: \textbf{TD3B}) that plays a key role in all biology.

\subsection{Agency, Self-Causation, and Causal Closure}\label{sec:agency_causal_closure}
Agency clearly takes place at the psychological level. People plan and, with greater or lesser success, carry out those plans \cite{Gray}, thus altering features of the physical world. In this way, technological developments such as farming and metallurgy and abstract ideas such as  the design of an aircraft or a digital computer have causal power \cite{Ellis_2016} and alter history \cite{Bronowski 2011}. 

The emergent psychological dynamics of the brain demonstrably has real causal powers.  So how does such agency occur?

\paragraph{Self-causing systems}
Agency is centrally related to the idea of a \textit{self-causing system}. The idea of a \textit{system} is crucial, ``an integration of parts into an orderly whole that functions as an organic unity'' (\cite{Juarrero 2002}:108-111).  This enables self-causation (\cite{Juarrero 2002}:252):
\begin{quote}
	``\textit{Complex adaptive systems exhibit true self-cause: parts interact to produce novel, emergent wholes; in turn these distributed wholes as wholes regulate and constrain the parts that make them up''.}
\end{quote}
(\cite{Murphy and Brown}:85-104) develop the theme further, emphasizing firstly how a complex adaptive system represents the emergence of a system with a capacity to control itself. Secondly,   
agency is related to the variation and selection process emphasized here: the dynamical organisation of a complex adaptive system functions as an internal selection process, established by the system itself, that operates top-down to preserve and enhance itself. 
This process is an example of the interlevel  
causal closure that is central to biology \cite{Mossio 2013} \cite{Ellis_2020b_Closure}. It leads to the circularity of the embodied mind \cite{Fuchs}:
\begin{quote}
	``\textit{From an embodied and enactive point of view, the mind-body problem has been reformulated as the relation between the lived or subject body on the one hand and the physiological or object body on the other. 
		The aim of the paper is to explore the concept of circularity as a means of explaining the relation between the phenomenology of lived experience and the dynamics of organism-environment interactions. 
		.. It will be developed in a threefold way:}
		
		(1) \textit{As the circular structure of embodiment, which manifests itself (a) in the homeostatic cycles between the brain and body and (b) in the sensorimotor cycles between the brain, body, and environment. This includes the interdependence of an organism's dispositions of sense-making and the affordances of the environment.
}		

		(2) \textit{As the circular causality, which characterizes the relation between parts and whole within the living organism as well as within the organism-environment system.}
		
		(3) \textit{As the circularity of process and structure in development and learning. Here, it will be argued that subjective experience constitutes a process of sense-making that implies (neuro-)physiological processes so as to form modified neuronal structures, which in turn enable altered future interactions.}
		
		\textit{On this basis, embodied experience may ultimately be conceived as the integration of brain-body and body-environment interactions, which has a top-down, formative, or ordering effect on physiological processes.''}
\end{quote}
This is also related to the 
Information Closure theory of consciousness \cite{Chang et al 2020}.:\textit{ ``We hypothesize that conscious processes are processes which form non-trivial informational closure (NTIC) with respect to the environment at certain coarse-grained scales. This hypothesis implies that conscious experience is confined due to informational closure from conscious processing to other coarse-grained scales.''}


\paragraph{The predictive coding view} Intentional action is a process of agent selection from various possibilities.  This possibility of agency is congruent with the predictive coding view, as three examples will demonstrate. 
Firstly \cite{Seth et al 2012} state,
\begin{quote}
	``\textit{We describe a theoretical model of the neurocognitive mechanisms underlying conscious presence and its disturbances. The model is based on interoceptive prediction error and is informed by predictive models of agency, general models of hierarchical predictive coding and dopamine signaling in cortex ...
		The model associates presence with successful suppression by top-down predictions of informative interoceptive signals evoked by autonomic control signals and, indirectly, by visceral responses to afferent sensory signals. 
		The model connects presence to agency by allowing that predicted interoceptive signals will depend on whether afferent sensory signals are determined, by a parallel predictive-coding mechanism, to be self-generated or externally caused.''}
\end{quote}

Secondly \cite{Negru} puts it this way:
\begin{quote}
	``\textit{The aim of this paper is to extend the discussion on the free-energy principle
		(FEP), from the predictive coding theory, which is an explanatory theory of the brain,
		to the problem of autonomy of self-organizing living systems. From the point of view
		of self-organization of living systems, FEP implies that biological organisms, due to the
		systemic coupling with the world, are characterized by an ongoing flow of exchanging
		information and energy with the environment, which has to be controlled in order to
		maintain the integrity of the organism. In terms of dynamical system theory, this means
		that living systems have a dynamic state space, which can be configured by the way
		they control the free-energy. In the process of controlling their free-energy and modeling
		of the state space, an important role is played by the anticipatory structures of the
		organisms, which would reduce the external surprises and adjust the behavior of the
		organism by anticipating the changes in the environment. In this way, in the dynamic
		state space of a living system new behavioral patterns emerge enabling new degrees of
		freedom at the level of the whole.''}
\end{quote}
Finally \cite{Szafron 2019} characterizes it thus:
\begin{quote}
	\textit{``
		Using the Free
		Energy Principle and Active Inference framework, I  describe a particular mechanism for
		intentional action selection via consciously imagined goal realization, where contrasts between
		desired and present states influence ongoing neural activity/policy selection via predictive coding
		mechanisms and backward-chained imaginings (as self-realizing predictions). A radically
		embodied developmental legacy suggests that these imaginings may be intentionally shaped by
		(internalized) partially-expressed motor predictions and mental simulations, so providing a
		means for agentic control of attention, working memory, and behavior.
		''}
\end{quote}
The overall result is a final higher level organising principle that acts as an attractor state during evolution:
\begin{description}
	\item[STB3] \textbf{Stochasticity and Agency} \textit{Stochasticity at the micro level allows macro level dynamics to select preferred micro outcomes, thus freeing higher levels from the tyranny of domination by lower levels. 
		By this mechanism, downward selection of preferred  micro outcomes enables self-causation and agency.}
\end{description}
The big picture is that randomness is rife in biology. Evolutionary processes have adapted biological systems to take advantage of this \cite{Hoffman}, with higher level processes selecting preferred outcomes from a variety of possibilities at the lower levels, thereby enabling the higher level organising principles characterised in the previous two sections to shape physical outcomes \cite{Noble and Noble 2018 Stochasticity}. The underlying physics enables this, but does not by itself determine the particular outcomes that occur, for they are contextually determined via time dependent dynamical constraints (\cite{Juarrero 2002}:131-162).

\section{The Whole Universe Gambit and Causal Closure}\label{sec:whole_universe}
The hardcore reductionist responds to the previous sections by saying yes the brain is an open system, but the universe as a whole  is not. Extend your micro data to include all the particles in the universe - well, in the region of the universe that is causally connected to us (i.e inside the particle horizon \cite{Hawking and Ellis 1973}) - and it is then  causally closed. All the data incoming to the brain is determined by causally complete microphysical processes in this extended domain, hence brain outcomes are determined by them too.  

The response, denying that this can work, has many levels. Please note that as stated before I am concerned with the possibility of physics determining specific outcomes, such as the words in Carlo's emails, not just statistical outcomes. His emails did not contain a statistical jumble of letters or words: they contained rational arguments stated coherently. This is what has to be explained.  The question is how the underlying physics relates to such specific rational outcomes. 

In order of  increasing practical importance the issues are as follows.

 First, Section \ref{sec:micor_determine} denies that the micro physical level is in fact causally complete, because of irreducible quantum indeterminism. While this can indeed have an effect on the brain, its primary importance is to deny that physics at the micro level is in principle causally complete.

Second, Section \ref{sec:mental states} makes the case that even if the incoming data was determined uniquely by microphysics everywhere in the surroundings, they would not determine a unique brain micro state in any individual  because 
of the 
multiple realisability of macro states by microstates  (\S\ref{sec:multiple_realise}).  

Third, Section \ref{sec:macro_determine} makes the case that important aspects of macro physics are in practice indeterminate because of the combination of chaotic dynamics and the impossibility of specifying initial data to infinite precision. This has neural outcomes \textit{inter alia} because it applies to weather patterns and forest fires.

Fourth, Section \ref{sec:Bio-Random} points out that there is considerable randomness in the external world biology that  the mind interacts with at both micro and macro levels.  These biological outcomes are not precisely predictable from their micro physics initial data. It has key impacts on the mind related in particular to the relations between humans and viruses.

Fifth, Section \ref{sec:social_brain_understand} points out that because the brain is a social brain (\S\ref{sec:social brain}), its macro level responses to incoming data are not purely mechanical: they are highly sophisticated responses at the psychological level to social interactions. These  are affected by unpredictable effects such as weather and pandemics. By the mechanisms discussed in Section \ref{sec:Plastic_brain}, these understandings reach down to structure the neural context of brain microphysics.
 
 Finally Section \ref{sec:causal_closure} makes the case that  the larger environment interacts with the brain  by providing the setting for interlevel circular causation. This can be claimed to be the real nature of causal closure. It is what is involved in order to have the data, constraints, and boundary conditions needed to determine specific outcomes in real world contexts, enabling self-causation. This is the opposite of being determined by microphysics alone. 


\subsection{Is micro physics causally complete?}\label{sec:micor_determine}

Carlo's argument is that micro data dependence of all outcomes undermines the possibility of strong emergence.  To summarise, suppose I am given the initial positions $\textbf{r}_i$ and momenta $\textbf{p}_i$ of all particles in the set 
${\cal P}$ everywhere,\footnote{``Everywhere'' means within the particle horizon \cite{Hawking and Ellis 1973}.} where 
\begin{equation}\label{eq:L1}
{\cal P} := \textrm{(protons, neutron, electrons)}
\end{equation}
at a foundational level \textbf{L1}. At a higher level \textbf{L2} this constitute an emergent structure \textbf{S}, such as a neural network. The details of \textbf{S} are determined by the microdata, even though its nature cannot be recognised or described at level \textbf{L1}. The forces between the particles at level \textbf{L1} completely determine the  dynamics at level \textbf{L1}. Hence the emergent outcomes at level \textbf{L2} are fully determined by the data at level \textbf{L1}, so the emergence of dynamical properties and outcomes at level \textbf{L2} must be weak emergence and be predictable, at least in principle, from the state (\ref{eq:L1}) of level \textbf{L1}, even if carrying out the relevant computations is not possible in practice.  This would apply equally to physical, engineering, and biological emergent systems. It is in effect a restatement of the argument from supervenience.\\

\noindent There are problems with the argument just stated as regards both microphysics and  macrophysics.

\paragraph{Quantum physics uncertainty relation} 
The  Heisenberg \href{https://en.wikipedia.org/wiki/Uncertainty_principle}{uncertainty relations} undermine this Laplacian dream because initial data cannot be specified with arbitrary accuracy \cite{Heisenberg}.
The standard deviations of position $\sigma_x$ and  momentum $\sigma_p$ obeys
\begin{equation}\label{eq:uncertainty}
\sigma_x \sigma_p \geq \hbar/2
\end{equation}
\cite{Kennard}, so one cannot in principle set initial data precisely at level \textbf{L1}. Consequently, outcomes based on standard Lagrangians dependent on $x$ and $p$ 
are uncertain in principle.

 Essentially the same issue arise in the case of classical physics \cite{Gisin_2019}  because  data cannot be prescribed to infinite accuracy \cite{Ellis Meissner Nicolai}. Further in the quasi-classical approximation, it will be subject to the uncertainty (\ref{eq:uncertainty}), reinforcing that conclusion  \cite{Del Santo}. This affects outcomes of chaotic dynamics (\S \ref{sec:macro_determine}). 

\paragraph{Irreducible uncertainty in quantum outcomes} There is irreducible uncertainty of quantum outcomes when wave function collapse to an eigenstate takes place, with outcomes only predictable statistically via the Born rule \cite{Ghirardi}. One cannot for example predict when an excited atom will emit a high energy photon, one can only predict the probability of such an event
This unpredictability has consequences that can get
amplified to macrolevels, for example causing errors in computer memories due to cosmic rays \cite{Cosmicray_compuer} \cite{Cosmicray_computer1}. The specific errors that occur are not determined by physics, because quantum physics is foundationally causally incomplete. 

 
 \paragraph{Biological damage due to cosmic rays}
 Cosmic rays can alter genes significantly enough to cause cancer. In particular, galactic cosmic rays lead to significant fatality risks for long-term space missions. 
 This is discussed in 
 \cite{cosmic rays1} \cite{cosmic rays2} \cite{cancer5}
 This shows both the contextual dependence of local outcomes in this case, and their unpredictability in principle. 
 
 \paragraph{Unpredictable brain effects}
 This obviously can affect the mental processes of those  undertaking space travel. 
 The brain 
 can be affected crucially by distant events that are in principle unpredictable because they result from quantum decay of excited atoms. 
 
 The statistics of outcomes  is strictly predicted by quantum theory. But in terms of causal completeness of biological events, we wish to know  which specific person gets affected at what specific time, thereby changing individual thought patterns. Detailed microphysical initial data everywhere cannot tell us that.
 
  This is a situation that only affects a small number of people, but it is important  because it establishes that \textit{in principle} the physicalist whole world gambit does not work (after all, that argument is an in principle argument: no one argues that it can work in practice in terms of allowing actual predictions of unique biological outcomes). 

\subsection{Mental states  and multiple realisability}\label{sec:mental states}
Incoming sensory data in a real world context affects brain macrostates 
which  then shape micro level connections via learning. 
But they do not do so in a unique way: 
incoming sensory data does determine unique brain microstates 
because of the multiple realisability of higher level states at the physical level.

\paragraph{Mental states and multiple realisability}
A given set of incoming data does not result in a unique brain physical microstate because of multiple realisability of the higher level state at the lower level (Section \ref{sec:multiple_realise}). 
This is a key property of brain function. \cite{Silberstein} state
\begin{quote}
	\textit{
		``Functionalists (and others) claim that mental states are radically multi-realizable, i.e., that mental states like believing that $p$ and desiring that $p$ can be multiply realized within individual across time, across individuals of the same species, across species themselves, and across physical kinds such as robots. If this is true, it raises crucial question: why do these states always have the same behavioural effects? In general we expect physically similar states to have similar effects and different ones to have different effects. So some explanation is required of why physically disparate systems produce the same behavioural effects. If there is nothing physically common to the `realizations' of a given mental state, then there is no possibility of any uniform naturalistic explanation of why the states give rise to the same mental and physical outcomes.'' 
	}
\end{quote}
The brain responds to incoming data via the predictive processing mechanisms discussed in Section \ref{sec:brain_predictive_mechanisms} , with updating of the relevant expectations taking place all the time on the basis of experience.  
The macro psychological processes that occur in this way  reach down to shape neural network connections and weights \cite{Kandel_memory} \cite{Kandel psych} in ways that are not unique. 
These then change constraints at the electron/ion level, realising any one of the billions of possible changes at that level that are in the right equivalence class. 

\paragraph{Unpredictable brain effects} Unique micro level physical conditions in the brain (the specific details of constraints in the electron/ion Lagrangian that will determine the ongoing brain dynamics) cannot in principle be determined by incoming data from the external world because of multiple realisability. Ordered outcomes appear at the brain macro level according to the predictive coding logic outlined in Section \ref{sec:stochastic_brain}, which then activates any one of the microstates in the corresponding equivalence class at the micro level (Section \ref{sec:multiple_realise}). 
All of this can of course be traced at the microphysical level, both internal to the brain and externally. But what is driving it is psychological level understandings.

\subsection{Is macro physics causally complete?}\label{sec:macro_determine}

The atmosphere is an open system dynamically driven by the Sun's radiation,  and with vary complex interactions taking place between the atmosphere, subject to winds and convection, water (the seas and lakes and clouds and ice), and land \cite{Ghil  and Lucarini 2020}. These are unpredictable in detail because of chaotic dynamics.

\paragraph{Convection patterns}
Consider a higher physical level \textbf{L3} in the context of a fluid where 
convection patterns take place. Because of the associated chaotic dynamics together with the impossibility (\ref{eq:uncertainty}) of setting initial data to infinite precision (Section \ref{sec:micor_determine}), macroscopic outcomes are  unpredictable in principle from micro data (\ref{eq:L1}) .  Convection patterns are an example  \cite{Bishop_2008_convection}: an extremely small perturbation in a fluid trapped between two levels  where a heat differential is maintained can influence the particular kind of convection that arises. 
 \cite{Anderson_01} puts it this way: 
\begin{quote}
	``\textit{A fluid dynamicist when studying the chaotic outcome of convection in a Benard cell knows to a gnat's eyelash the equations of motion of this fluid but also knows, through the operation of those equations of motion, that the details of the outcome are fundamentally unpredictable, although he hopes to get to understand the gross behaviour. This aspect is an example of a very general reality: the existence of universal law does not, in general, produce deterministic, cause-and-effect behaviour.}''
\end{quote}
The outcome is an emergent layer of unpredictability at both local scales (thunderstorms, tornados, typhoons,  and so on) and globally (large scale weather outcomes). The latter are famously characterised by strange attractors  \cite{Lorenz 1963}, involving instability and fractals, but much more importantly interactions between different length scales that make prediction  impossible in principle \cite{Lorenz 1969}, as discussed in depth by \cite{Palmer}.
Downward constraints then entrains lower level dynamics to follow, as stated by  \cite{Bishop 2012}:  ``\textit{Large-scale structures arise out of
		fluid molecules, but they also dynamically condition
		or constrain the contributions the fluid molecules can
		make, namely by modifying or selecting which states
		of motion are accessible to the fluid molecules''}. %

\paragraph{The Butterfly Effect}
 Lorenz intended the phrase `the butterfly effect' to describe the existence of
an absolute finite-time predictability barrier in certain multi-scale fluid systems,
implying a breakdown of continuous dependence on initial conditions for large
enough forecast lead times \cite{Palmer}. \cite{Lorenz 1969}  states \begin{quote}
	``\textit{It is proposed that certain formally deterministic fluid systems which possess many
scales of motion are observationally indistinguishable from indeterministic systems. Specifically, that two states of the system differing initially by a small observational
error will evolve into two states differing as greatly as randomly chosen states of
the system within a finite time interval, which cannot be lengthened by reducing the
amplitude of the initial error}.''
\end{quote} This happens because of the interactions between the different length scales involved. Palmer's illuminating paper \cite{Palmer} concludes that this real butterfly effect\footnote{See \href{https://www.youtube.com/watch?v=vkQEqXAz44I}{https://www.youtube.com/watch?v=vkQEqXAz44I} for an enlightening lecture on the common and real butterfly effects. The implication is that you need ensemble forecasts.} does indeed occur - but only for some particular sets of  initial data. Nevertheless occurring from time to time denies causal closure of physics on this scale in practice. That clearly means the underlying physics at the particle level cannot have been causally closed either. The multiscale weather dynamics studied by \cite{Lorenz 1969} reaches down to influence atomic and electron motions (think thunderstorms) at the lower physics level. But this is the crucial point: you cannot predict when those cases will occur.

\paragraph{Forest Fires}
are an example of self-organised critical behaviour \cite{Malamud} affected by local atmospheric convection activity in that firstly many forest fires are cased by lightning\footnote{\href{https://www.nytimes.com/2020/08/23/us/dry-thunderstorms-california-fires.html?campaign_id=51&emc=edit_MBE_p_20200824&instance_id=21553&nl=morning-briefing&regi_id=92041267&section=whatElse&segment_id=36797&te=1&user_id=5c8c3cd2866b46cd9e89b9944bb1deea}{Dry Thunderstorms Could Accelerate the California Wildfires}, 
}, and secondly the spread of the fire is determined by local winds which are changed by local convection effects due to the fire. The detailed dynamics of the fire are unpredictable because of these 
links; even probabilities are tricky \cite{Mata}.

\paragraph{Unpredictable brain effects}
In terms of the effect on the brain, these random outcomes shape decisions from whether to open an umbrella on a trip to the shops, to farmers' decisions as to when to harvest crops, aircraft pilots decisions about \textit{en route} flight planning, and homeowners decisions about whether or not to flee a forest fire. It causes an essential unpredictability in mental outcomes. This is a first reason the external world has an ongoing unpredictable key effect on individual brains. 


\subsection{Biological Randomness: the Microbiome}\label{sec:Bio-Random}
Biological dynamics in the external world are subject to unavoidable uncertainty because of the random nature of molecular level events, already alluded to in Section \ref{sec:bio_stochastic}.

\paragraph{Interacting microbiomes and viruses} The immense complexity of each individual person's microbiome (Section \ref{sec:bio_stochastic})  interacts, through social events, with other people's microbiomes, as do their viruses. 
 Genetic variability is central to the mutation  of  microbes and viruses 
  in the external world. Detailed physical microstates everywhere determine the statistics of such variations, but not the specific ones that  actually occur, which are due \textit{inter alia} to mutation and  recombination and horizontal gene transfer in the case of microbes, mistakes by RNA or DNA polymerases, radiation- or chemical- or host cell defences-induced mutation, and re-assortment in the case of viruses.  
 Predicting mutations is essentially impossible, even for viruses with $10,000$ bases like HIV. All you CAN say is that the known mutation rate for that organism 
  predicts that every single copy of the HIV genome (for example) will have at least one mutation ($10^{-4}$ rate).\footnote{I thank Ed Rybicki for this comment.} 
\paragraph{Unpredictable brain effects}
This has crucial effects in our brains 
 that are completely unpredictable because firstly of the randomness of the genetic mutations leading to these specific microbes and viruses, and secondly of the details of the events that lead to their spread through animal and human populations; this can all be expressed in terms of rugged adaptive landscapes \cite{Orr 2005}. This firstly directly  affects human health and brain dynamics in each of the set of interacting  brains (\S \ref{sec:social_brain_understand}) via the gut brain axis \cite{Cryan}, and 
 then plays a key role in individual   associated mental events   
 such as individual planning of what do to about anxiety,
 obesity, schizophrenia, Parkinson's disease, and Alzheimer's disease, or flu, AIDS or  COVID-19. This is a second reason the external world has an important unpredictable key effect on individual brains.

\subsection{Social understandings and individual brains}\label{sec:social_brain_understand}
Our brain is a social brain (Section \ref{sec:social brain}). 
Information from the external world affects mental states via ongoing complex social interactions, which have real causal powers. They structure our mental activities in everyday life.

\paragraph{Social understandings}
There is an intricate relation between the individual and society \cite{Berger}  \cite{Longres} \cite{Berger and Luckmann 1991} \cite{Donald 2001}  and between  individuals and institutions in a society \cite{Vass} \cite{Vass 2012}. 
The downward effect of the social context on an individual brain is mediated by social interactions and understandings (Section \ref{sec:social brain}).
In this social context, a complex interaction  takes place involving mind reading, prediction, filling in of missing data, taking social context into account  \cite{Longres} 
\cite{Donald 2001} \cite{Frith 2009}. 
This nature of the interactions of a many brains, each a self causing open system (Section \ref{sec:agency_causal_closure}), is the main practical day to day reason that microphysics everywhere cannot determine unique outcomes in each of the brains involved.  Downward causation from the social level interactions to individual brains  to the underlying molecular biology and thence physical levels is the causal chain. 

\paragraph{Abstract variables have causal powers}
This is all enabled by our symbolic ability \cite{Deacon 1997}, resulting in our use of spoken and written language, which is the key factor enabling this to happen 
\cite{Ginsburg and Jablonka 2019}  (Section \ref{sec:symbolic brain}). This affects our individual brain operations as we consider the continually changing detailed implications of money, closed corporations, laws, passports, and so on in our lives. 

\paragraph{Policy decisions have causal powers}
Given this context, social variables have causal power \cite{Longres} \cite{Harari 2014} and affect brain states; in particular, this applies to policy decisions. 
The interaction outcomes  are shaped at the social level, which is where the real causal power resides, and then affect individual brain states in a 
downward way. Complex interpretative processes take place shaping psychological level reactions, which then shape neural network and synapse level changes in  a contextual way on an ongoing basis 
as studied by social neuroscience  \cite{Cacciopo}.

\paragraph{Unpredictable brain effects}
Policy decisions are sometimes based in unpredictable events such as cyclones or forest fires 
(Section \ref{sec:macro_determine}) or a global pandemic or local infectious outbreak  (Section \ref{sec:Bio-Random}). 
Mandatory evacuating of towns in the face of a cyclone or wild fire,\footnote{For a typical context see \href{https://www.nytimes.com/2020/08/20/us/ca-fires.html}{https://www.nytimes.com/2020/08/20/us/ca-fires.html}.} going into shelters in the case of a tornado, or policy decisions such as lockdowns in the face of a pandemic will all be unpredictable because their cause is unpredictable, and so will cause unpredictable outcomes in individual brains at macro and micro levels. The causal chain is an unpredictable trigger event, followed by a social policy choice that then changes outcomes in individual brains.   Detailed physical data everywhere enables this to happen by providing the basis for stochastic outcomes that cannot be determined uniquely from that data be because of the real butterfly effect in the case of weather, and its analogue in the case of microbe and viral mutations.  Carlo's view that microphysics determines all brain dynamics in this extended context could hold if it were not for the random nature of the trigger events. 

\subsection{Real Causal Closure}\label{sec:causal_closure}
Carlo's move of bringing into focus the larger context  is certainly correct in the following sense: the way that causal closure takes place in reality  involves the whole environment. But that means it is an interlevel affair, for the environment involves all scales.  


\paragraph{The real nature of causal closure}
\begin{itemize}
	\item From my viewpoint, what is meant by the phrase ``causal closure'' as used by Carlo and other physicists is in fact that one is talking \textit{about existence of a well-posed effective theory} $\textbf{EF}_\textbf{L}$ that holds at some emergent level \textbf{L}. 
This means data $d_\textbf{L}$ for variables $v_\textbf{L}$ at that level \textbf{L} specifies unique outcomes, or the statistics of such outcomes, at that level.

	\item However existence of such a theory does  by itself not determine any specific physical outcomes. It implies that \textit{if} the right data and boundary conditions are present, \textit{and} all constraints that hold are specified, \textit{then} a unique or statistical outcome is predicted by the physics at that level.
	
	\item It does not attempt to say where that data, boundary conditions, and constraints come from. But without them you do not have causal closure in what should be taken to be the real meaning of the term: \textit{sufficient conditions are present to causally determine real world outcomes that happen.} For example social dynamics are active causal factors that reach down to affect physics outcomes, as is abundantly clear in the COVID-19 crisis:  policies about face masks affect physical outcomes.

\item My use of the term, as developed in full in \cite{Ellis_2020b_Closure}, regards causal closure as interlevel affair,  such as is vital to biological emergence \cite{Mossio 2013}. The conjunction of upward and downward effects must self-consistently determine the boundary conditions, constraints, and initial data at a sufficient set of levels that unique or statistical outcomes are in fact determined by the interlocking whole.

\item When that happens you can of course trace what is happening at whatever physics level you choose as a base level \textbf{L0}. But over time, the later initial data, boundary conditions, and constraints at that level are dynamically affected by the downward mechanisms outlined in Section \ref{sec:causation_types}. Because causation is equally real at each level, the higher levels are just as much key factors in the causal nexus as is level \textbf{L0}, as time proceeds.   Higher Level Organising Principles, independent of the lower level physics, shape physical outcomes.
The state of variables at level \textbf{L0} at time $t_0$ uniquely determines the higher levels at that time, but not at a later time $t_1>t_0$.

\item The freedom for higher levels to select preferred lower level outcomes exists because of the stochastic nature of biological processes at the cellular level (Section \ref{sec:stochastic_select}). 
 
\item The illusion of the effective theory at a physical level \textbf{L0} being causally complete is because physicists neglect to take into account their own role in the experiments that establish the validity of the effective theory that holds at that level. When you take that role into account, those experiments involve causal closure of all levels from \textbf{L0} to the psychological level \textbf{L6} where experiments are planned and the social level \textbf{L7} which enables the experimental apparatus to come into being. 
\end{itemize}
Another term used for causal closure in this sense is \textit{operational closure}: the organisational form of the processes that enable autopoietic self-production and conservation of system boundaries  \cite{Di Paolo and Thompso 2014} \cite{Ramstaed et al 2019}.


\paragraph{The predictive coding/free energy viewpoint}
My view agrees with the growing predictive coding consensus, as presented in previous sections. Karl Friston (private communication) says the following: 
\begin{quote}
	``\textit{I imagine that downward causation is an integral part of the free energy formalism; particularly, its predication on Markov blankets.
		I say this in the sense that I have grown up with a commitment to the circular causality implicit in synergetics and the slaving principle (c.f., the Centre Manifold Theorem in dynamical systems). As such, it would be difficult to articulate any mechanics without the downward causation which completes the requisite circular causality. Practically, this becomes explicit when deriving a renormalisation group for Markov blankets. We use exactly the same formalism that Herman Haken uses in his treatments of the slaving principle \cite{Haken} \cite{Haken and Wunderlin} to show that Markov blankets of Markov blankets constitute a renormalisation group. If existence entails a Markov blanket, then downward causation (in the sense of the slaving principle) must be an existential imperative.
''}
\end{quote}
The final conclusion of this section is the following
\begin{quote}
	\textbf{Unique causal outcomes in individual microphysical brain states   do not occur when one includes 
		causal effects of the external world.   \textit{This does not work (i) because  microphysics is not in fact causally closed due to quantum wave function collapse,  
		(ii)  external information cannot uniquely determine microphysical states in the brain - multiple realisability makes this impossible, (iii) unpredictable macro level chaotic dynamics occurs, (iv) microbiome dynamics that affect brain states is unpredictable, and (v) the way external states influence brain states is strongly socially determined and can include events that are in principle unpredictable.
		 }
}\end{quote}
However it certainly is true that such downward causal effects on individual brains occur. They just do not do so in a way that is uniquely determined by physical effects alone.

\section{Microphysics Enables but Does Not Determine}\label{sec:conclude}
In this section I summarise my response (Section \ref{sec:response_summary}),  and comment on the relation of all the above to the issue of free will (Section \ref{sec:FreeWill}) and to the possibility spaces that are the deep structure of the cosmos (Section \ref{sec:possible}).

\subsection{The basic response}\label{sec:response_summary}
There are a series of key issues that shape my response. 
\begin{itemize}
	\item I am concerned with what determines the specific outcomes that occur in real world contexts, not just with statistical prediction of outcomes. How does physics underlie the existence of a Ming dynasty teapot? Of the particular digital computer on which I am typing this response? Of Einstein's publication of his paper on General Relativity? Of the election of Donald Trump as President of the United States of America?  
\item Consider a specific individual brain at a particular time. The difference between synchronic and diachronic emergence is key. Carlo's view can be defended in the synchronic emergence case, but cannot be correct in the diachronic case, because individual brains are open systems. The initial microphysical state of the brain simply does not include all the data that determine its outcomes at later times. 
\end{itemize}
This is what is discussed in depth in Sections \ref{sec:brain_open_system} - \ref{sec:stochasticity}.

\paragraph{The whole universe context}
 Claiming that this problem is solved by going to a larger scale where causal closure does indeed hold (the cosmological scale), which therefore implies that the specific evolution of all subsystems such as individual brains is also uniquely determined, does not work for a series of reasons. 
 I list them now with the theoretically most important issues first. This is the inverse of the order that matters in terms of determining outcomes in practical terms. As far as that is concerned, the most important issues are the later ones. 
 
 It does not work because of,
\begin{enumerate}
	\item  Irreducible quantum uncertainty at the micro level which affects macro outcomes; this demonstrates that the claim is wrong in principle. It can indeed have macro effects on the brain, but this is not so important at present times because of the shielding effect of the earth's atmosphere. However it has played important role in evolutionary history  \cite{Percival},  as discussed by  \cite{cosmic rays3} \cite{Scalo_cosmicrays}
	\item The fact that downward effects from that larger context to the brain, which certainly occur  via neural plasticity and learning, cannot in principle determine a unique brain microstate, because of multiple realisability of those detailed physical states when this occurs. Unique brain microstates cannot occur in this way.  
	\item Uncertainty in principle at the ecosystem level due to chaotic dynamics and the real butterfly effect plus the inability to set initial molecular conditions precisely. This has major unpredictable effects on individual brains due to forest fires, thunderstorms, tornadoes, and tropical cyclones.   
	\item Microbiome dynamics that is in principle  unpredictable because of the molecular storm and huge number of molecules involved, plus the inability to set initial molecular conditions precisely. This affects  individual and social outcomes as evolution takes place on a rugged adaptive landscape that keeps changing as all the interacting species evolve. This crucially affects brain dynamics through the gut-brain axis.
	\item Social understandings that shape how  external signals are interpreted by the brain, when social level policies and choices (which may involve unpredictable events such as thunderstorm details or pandemic outbreaks) chain down to influence flows of electrons in axons. It simply is not a purely physics interaction.  
\end{enumerate}
Carlo's vision of the external world as a whole evolving uniquely and thereby determining unique brain states because they are a part of the whole, may work in some contexts where irreducible uncertainty 1. and effective uncertainty 3. and 4. do not occur. It cannot however work when any of these effects come into play, which certainly happens in the real world. This demonstrates that as a matter of principle, it is the higher level effects - psychological and social variables - that are sometimes calling the tune. But that means they are always effectively doing so in the social context which is the habitat of minds.  

\paragraph{Causal closure}

 Real world causal closure is an interlevel affair, with microphysical outcomes determined by features ranging from global warming and tropical cyclones to COVID-19 policy decisions. It simply cannot occur at the microphysics level alone. Some of the effective variables which have changed human history are abstract concepts such as the invention of arithmetic, the concept of money, and the idea of a closed corporation.  These have all crucially affected microphysical outcomes, as have abstract theories such as the theory of the laser  and the concepts of algorithms, compilers, and the internet.\\

\noindent Overall, as stated by \cite{Bishop 2012}, the situation is that 
\begin{quote}
\textit{``Whatever necessity the laws of physics
	carry, it is always conditioned by the context into
	which the laws come to expression.''
}\end{quote}
So in response to Carlo's final email (Section \ref{sec:dialogue}):
\begin{description}
	\item[CR] Today the burden of the proof is not on this side.  Is on the opposite side.  Because:
	
	\textbf{(i)}  There is no single phenomenon in the world where microphysics has been proven wrong (in its domain of validity of velocity, energy, size...).
	
	\item[GE] The view I put respects the microphysics completely. Of course it underlies all emergent phenomena, without exception. Microphysics certainly is not wrong.
	
\item[CR]	\textbf{(ii)}  By induction and Occam's razor, is a good assumption that in its domain validity it holds.
	
	\item[GE] Yes it holds in its domain of validity, which is at the microscale. The question at stake is, How much larger is its domain of validity?  I think the comment is meant to say that its domain of validity includes biology and the brain, in the sense that, by itself it fully determines all biological and brain outcomes.
	
	 However completely new kinds of behaviour emerge in the biological domain. The kind of causation that emerges is simply different than the kind of statistically determinist relation  between data and outcomes that holds at the microphysical level. The microphysics allows this emergence: it lies within the space of possibilities determined by that physics. In that sense the higher level outcomes lie within the domain of validity of the microphysics. But the microphysics by itself does not determine the macro level outcomes (see the listed points above). Occam's razor does not work.
	
\item[CR]	\textbf{(iii)} There are phenomena too complex to calculate explicitly with microphysics.  These provide no evidence against (ii), they only testify to our limited tools.
	
	\item[GE] Carlo only considers efficient causation, because that is what physicists study.  As Aristotle pointed out \cite{Bodnar 2018}, that is only one of the four kinds of causation that occur in the real world (Section \ref{sec:causation_types}). In the real world the other kinds of causation play a key role in determining outcomes. All four are needed to determine specific outcomes.  
\end{description}	
	I accept the need to provide the burden of proof. I have done so in the preceding sections. 




\subsection{What about Free Will?}\label{sec:FreeWill}
The implication of Carlo's argument is that the causal power of microphysics prevents the existence of free will. This touches on a vast and complex debate. My arguments above deny that the underlying physics can disprove existence of free will in a meaningful sense. But that does not dispose of the debate. Does neurobiology/neuroscience deny free will?

\paragraph{Incomplete reductionism: neurobiology}
Francis Crick  gives a neuroscience based reductionist argument regarding the brain in  \href{https://en.wikipedia.org/wiki/The_Astonishing_Hypothesis}{\textit{The astonishing hypothesis}} \cite{Crick 1994}:
\begin{quote} 
	\textit{``You, your joys and your sorrows, your memories and your ambitions, your sense of personal identity and free will, are in fact no more than the behavior of a vast assembly of nerve cells and their associated molecules. ''} 
\end{quote}
Now the interesting point is that this a denial of Carlo's arguments. Crick is assuming that the real level of causation is at the cellular and molecular biology levels: that is where the action is, it is at that level that physical outcomes are determined. The implication is that this is what determines what specific dynamical outcomes take place at the underlying physical level - which is my position \cite{Ellis_Kopel_2019}. 

\paragraph{Free will and neurobiology} As to free will itself, does neurobiology and neuroscience deny its existence? That is a long and fraught debate related to intentionality and agency. Amongst the deeply considered books that argue for meaningful free will are \cite{Donald 2001} \cite{Dupre} \cite{Murphy and Brown} \cite{Murphy Oconnor Ellis}  \cite{Baggini 2015}. \cite{Frith 2013} has a nuanced discussion on agency and free will. \cite{Murphy and Brown} conclude (page 305) that ``\textit{free will should be understood as being the primary cause of one's own actions; this is a holistic capacity of mature, self-reflective human organisms acting within suitable social contexts}''.  This is essentially the consensus of the authors just named, Libet's experiments notwithstanding. Chris Frith \footnote{One of the topmost cited neuroscientists in the world: he had 203,843 citations on 2020/08/01. } expresses it this way \cite{Frith 2009}:
\begin{quote}
	`` \textit{I suggest that the physiological basis of free will, the spontaneous and intrinsic selection of one action rather than another, might be identified with mechanisms of top-down control. Top-down control is needed when, rather than responding to the most salient stimulus, we concentrate on the stimuli and actions relevant to the task we have chosen to perform. Top-down control is particularly relevant when we make our own decisions rather than following the instructions of an experimenter. Cognitive neuroscientists have studied top-down control extensively and have demonstrated an important role for dorsolateral prefrontal cortex and anterior cingulate cortex. If we consider the individual in isolation, then these regions are the likely location of will in the brain. However, individuals do not typically operate in isolation. The demonstration of will even in the simplest laboratory task depends upon an implicit agreement between the subject of the experiment and the experimenter. The top of top-down control is not to be found in the individual brain, but in the culture that is the human brain's environmental niche}''	
\end{quote}
This is a good description of both topdown effects in the brain \cite{Ellis 2018} and interlevel causal closure  \cite{Ellis_2020b_Closure}. It is also expressed well by \cite{Baggini 2015}.

\paragraph{Free will denialists don't really believe it} The physicist Anton Zeilinger told me the following story. He was once being harassed by someone who strongly argued that we do not have free will. Anton eventually in frustration reached out and slapped him in the face. He indignantly shouted, ``\textit{Why did you do that?}'', to which Anton responded ``\textit{Why do you ask me that question? You have just been explaining to me at length that I am not responsible for my actions. According to you, it's not a legitimate question.''}

If you have an academic theory about the nature of causation and free will, it must apply in real life too, not just when you are engaged in academic argumentation. If not, there is no reason whatever for anyone else to take it seriously - for you yourself do not.

\paragraph{Free will and the possibility of science} The ultimate point is that if we don't have meaningful free will, in the sense of the possibility of making rational choices between different possible understandings on the basis of coherence and evidence, then science as an enterprise is impossible. You then cannot in a meaningful way be responsible for assessing theories anyone proposes. The theory that free will does not exist causes the demise of any process of scientific investigation that is alleged to lead to that theory. We had better find a better theory - such as those in the books cited above.

  \paragraph{Denial of consciousness or qualia} Finally one should note that many pursuing the view that free will does not exists  also deny that consciousness and/or qualia exist and play any role in brain function. But neuroscience simply does not know how to solve the hard problem of consciousness \cite{Chalmers_1995}. As stated in \cite{Tallis 2016}, neuroscience helps define the necessary conditions for the existence of human consciousness and behaviour, but not the sufficient conditions. The self-defeating philosophical move of denying that consciousness and/or qualia exist (``what one canot explain does not exist'' \cite{Tallis 2016}) does not succeed in explanatory terms: how can you deny something if you have no consciousness? In that case, you do not satisfy the necessary conditions to deny anything: you do not exist in any meaningful sense \cite{Donald 2001}. For more on this see also  \cite{Gabriel 2017} \cite{Dennett and Strawson}.\\

\noindent  But in any event this is a different debate than my debate with Carlo, which in the end is about physics denying free will. 

\subsection{Possibility Spaces}\label{sec:possible}
All the biological effects discussed here are allowed by the underlying physical levels: they do not violate or alter those generic equations, which apply to all physical interactions without exception. 

A useful way to characterise this is in terms of \textit{possibility spaces}. In the case of physics, these  include phase spaces \cite{Arnold} (classical physics) and Hilbert spaces (quantum physics) \cite{Isham}. These are determined by the laws of physics, and are indeed equivalent to them: the laws allow the possibilities described by the possibility spaces, and the possibility spaces characterise the nature of the underlying laws.

Now the interesting point is that there are also biological possibility spaces. At the microbiology level they include a space of all possible proteins allowed by the laws of physics \cite{Wagner 2014}, of which only some have been realised on Earth by evolutionary processes \cite{Petsko_Ringe_2009}. Similarly there are sets of possible genotype-phenotype maps for metabolism and for gene regulation \cite{Wagner 2014}. 
There are limitations on physiological possibilities due to the nature of physics \cite{Vogel 2000} and consequent  scaling laws for biology \cite{West}. These are all immutable biological possibilities that, just like the laws of physics, are the same everywhere in the universe at all times, and allow life as we know it to come into existence in suitable habitats. \\

The claim one can make then is that Higher Level Organising principles such as those that are identified in this paper for biology in general and for the brain in particular  are also of this nature: they are timeless and eternal principles that can be expected to apply to life everywhere, because this is the only way it can work in principle. Thus these would include the possibility of metabolic networks,  genes and gene regulatory networks \cite{Wagner 2014}; of  physiological systems \cite{Physiology_human} \cite{Guyton 2016} and  developmental processes \cite{Wolpert}; of Evo-Devo type evolution  \cite{Carroll 2008}; and of social brains \cite{Dunbar 1998} operating on hierarchical predictive coding principles \cite{Clark 2013}, that are capable of analytical thought based in a symbolic capacity \cite{Deacon 1997}. \\

These are all possible, as they do indeed exist, 
 so the possibility of their existence is written into the nature of things. This is allowed by the underlying physics, but they represent higher order principles allowing life to come into existence and flourish, as discussed in depth by Stuart Kauffman inhos book \textit{At Home in the Universe} \cite{Kauffman at home}. 

\paragraph{Acknowledgments}  
I thank Carlo Rovelli for his patient dialogue with me regarding this issue. It is 
 a pleasant contrast with the arrogant dismissive comments and \textit{ad hominem} contemptuous personal attacks that are common in some reductionist circles and writings.

 I thank Karl Friston, Tim Palmer, and Ed Rybicki for helpful comments. 



\noindent Version 2020/09/06


\end{document}